\documentclass{article}
\usepackage{amsmath,amssymb,graphicx,amsfonts,epsf}
\usepackage{algorithmic}
\usepackage{algorithm}
\usepackage{tikz,xcolor}
\usepackage{graphicx} 
\usetikzlibrary{backgrounds,decorations.pathmorphing,arrows}

\def\EE{\mathbb E}

\def\RR{\mathbb R}

\def\ZZ{\mathbb Z}

\def\cG{\mathcal G}

\def\cI{\mathcal I}

\def\cP{\mathcal P}

\def\cR{\mathcal R}
\def\cM{\mathcal M}
\def\cN{\mathcal N}

\def\bx{\mathbf x}

\def\bx{\mathbf x}
\def\by{\mathbf y}

\def\bp{\mathbf p}

\def\b1{\mathbf 1}

\def\gd{\delta}

\def\eps{\varepsilon}

\def\cI{{\mathcal I}}

\newcommand{\qed}{\hfill$\square$\bigskip}
\newcommand{\raf}[1]{(\ref{#1})}
\newcommand{\proof}{\noindent {\bf Proof}~~}

\newcommand{\poly}{\operatorname{poly}}
\newcommand{\polylog}{\operatorname{polylog}}

\newcommand{\Max}{\textsc{White}}
\newcommand{\Min}{\textsc{Black}}
\newcommand{\Cl}{\operatorname{cl}}

\newcommand{\rank}{\operatorname{rank}}
\newcommand{\EXT}{\operatorname{EXT}}
\newcommand{\hide}[1]{}

\newcommand{\bone}{\ensuremath{\boldsymbol{1}}}

\newtheorem{theorem}{Theorem}

\newtheorem{lemma}{Lemma}
\newtheorem{corollary}{Corollary}
\newtheorem{proposition}{Proposition}
\newtheorem{observation}{Observation}

\newtheorem{remark}{Remark}

\newtheorem{Claim}{Claim}

\title{A Pseudo-Polynomial Algorithm for Mean Payoff Stochastic Games with Perfect Information and Few Random Positions\thanks{This paper combines the results announced in  preliminary versions in \cite{IPCO-2010,BEGM-ICALP13}. Part of this research was done at the Mathematisches Forschungsinstitut Oberwolfach during a stay within the "Research in Pairs" Program from July 26, 2015-August 15, 2015.
		The research of the third author has been partially funded by the Russian Academic Excellence Project '5-100'.}}
\author{
	Endre Boros\thanks {RUTCOR, Rutgers University, 640  Bartholomew
		Road, Piscataway NJ  08854-8003;
		(boros@rutcor.rutgers.edu)}
	\and
	Khaled Elbassioni\thanks {Masdar Institute of Science and Technology, Abu Dhabi, UAE;
		(kelbassioni@masdar.ac.ae)}
	\and
	Vladimir Gurvich\thanks {RUTCOR and MSIS, RBS, Rutgers University,
		100 Rockafeller Road, Piscataway, NJ 08854; and Dep. of Computer Sciences, National Research University Higher School of Economics (HSE), Moscow, Russia;
		(gurvich@rutcor.rutgers.edu)}
	\and
	Kazuhisa Makino\thanks{Research Institute for Mathematical Sciences (RIMS) 
		Kyoto University, Kyoto 606-8502, Japan;
		(makino@kurims.kyoto-u.ac.jp)}
}

\begin{document}
	\date{}
	\maketitle
	\begin{abstract}
		We consider two-person zero-sum stochastic mean payoff games
		with perfect information, or BWR-games, given by a digraph
		$G = (V, E)$, with local rewards $r: E \to \ZZ$, and
		three types of positions: black $V_B$, white $V_W$, and random $V_R$ forming a partition of $V$.
		It is a long-standing open question whether a polynomial time algorithm for BWR-games exists, or not, even when $|V_R|=0$.
		In fact, a pseudo-polynomial algorithm  for BWR-games would already imply their polynomial solvability.
		In this paper, we show that BWR-games with a constant number of random positions can be solved in pseudo-polynomial time.
		More precisely, in any BWR-game with $|V_R|=O(1)$, a saddle point in uniformly optimal pure stationary strategies can be found in time polynomial in $|V_W|+|V_B|$, the maximum absolute local reward, and the common denominator of the transition probabilities.
	\end{abstract}

	\section{Introduction}
	\label{intro}
	\subsection{Basic concepts}
	We consider two-person zero-sum
	stochastic games with perfect information and mean payoff:
	Let  $G = (V, E)$ be a digraph whose vertex-set  $V$
	is partitioned into three subsets  $V = V_B \cup V_W \cup V_R$
	that correspond to black, white, and random positions, controlled
	respectively, by two players, \Min\ - the \emph{minimizer}  and
	\Max\ - the \emph{maximizer}, and by nature.
	We also fix a {\em local reward} function  $r: E \to \ZZ$, and probabilities
	$p(v,u)$ for all arcs  $(v,u)$ going out of  $v \in V_R$.
	Vertices  $v \in V$  and  arcs $e \in E$  are called
	{\em positions} and {\em moves}, respectively.
	The game begins at time $t=0$ in the initial position $s_0=v_0$. In a general step, in time $t$, we are at position $s_t\in V$.
	The player who controls $s_t$ chooses an outgoing arc $e_{t+1}=(s_t,v)\in E$, and the game moves to position $s_{t+1}=v$. If $s_t\in V_R$ then an outgoing arc is chosen with the given probability $p(s_t,s_{t+1})$.  
	We assume, in fact without any loss of generality, that every vertex in $G$ has an outgoing arc.
	(Indeed, if not, one can add loops to terminal vertices.)
	In general, the strategy of the player is a policy by which (s)he chooses the outgoing arcs from the vertices (s)he controls. This policy may involve the knowledge of the previous steps as well as probabilistic decisions. We call a strategy {\it stationary} if it does not depend on the history and {\it pure} if it does not involve probabilistic decisions. For this type of games, it will be enough to consider only such strategies, since these games are known to be (polynomially) equivalent \cite{DGAA13} to the perfect information stochastic games considered by Gillette \cite{Gil57,LL69}. 
	
	In the course of this game players generate an infinite sequence
	of edges $\bp=(e_1,e_2,\ldots)$ (a \emph{play}) and the corresponding integer sequence
	$r(\bp)=(r(e_1),r(e_2),\ldots)$ of local rewards.
	At the end (after infinitely many steps) \Min\ pays \Max\ $\phi(r(\bp))$ amount.
	Naturally, \Max's aim is to create a play which maximizes $\phi(r(\bp))$, while \Min\ tries to minimize it.
	(Let us note that
	the local reward function $r:E\rightarrow \ZZ$ may have negative values, and
	$\phi(r(\bp))$ may also be negative, in which case
	\Max\ has to pay \Min. Let us also note that $r(\bp)$ is a random variable since random transitions occur at positions in $V_R$.) 
	Here $\phi$ stands for the \emph{limiting mean payoff}
	\begin{equation}\label{e0}
	\phi(r(\bp))=\liminf_{T\rightarrow\infty}\frac{\sum_{i=0}^T\EE[r(e_i)]}{T+1},
	\end{equation}
	where $\EE[r(e_i)]$ is the expected reward incurred at step $i$ of the play.
	
	As usual, a pair of (not necessarily pure or stationary) strategies is a saddle point if neither of the players can improve individually by changing her/his strategy. The corresponding $\phi(r(\bp))$ is the value of the game with respect to initial position $v_0$. Such a pair of strategies are called {\it optimal}; furthermore, it is called {\it uniformly optimal} if it provides the value of the game for any initial position.
	
	\medskip
	
	This class of \emph{BWR-games} was introduced in \cite{GKK88}; see also \cite{CH08}.
	The special case when $V_R = \emptyset$, \emph{BW-games}, is also known as
	{\em cyclic} games.
	They were introduced for the complete bipartite digraphs in \cite{Mou76a},
	for all (not necessarily complete) bipartite digraphs in \cite{EM79}, and for arbitrary digraphs\footnote{In fact, BW-games on arbitrary digraphs can be polynomially reduced to BW-games on bipartite graphs \cite{DGAA13}; moreover, the latter class can further be reduced to BW-games on complete bipartite graphs \cite{Vienna}.} in \cite{GKK88}.
	A more special case was considered extensively in the literature under
	the name of \emph{parity games} \cite{BV01,BV01a,CJH04}, \cite{Hal07,J98,JPZ06}, and later generalized
	also to include random positions in \cite{CH08}.
	A BWR-game is reduced to a \emph{minimum mean cycle problem} in case $V_W = V_R = \emptyset$,
	see, e.g., \cite{Kar78}.
	If one of the sets $V_B$ or $V_W$ is empty, we obtain
	a {\em Markov decision process} (MDP), which can be expressed as a linear program; see, e.g., \cite{MO70}.
	Finally, if both are empty $V_B = V_W = \emptyset$, we get a {\em weighted Markov chain}.
	In the special case when all rewards are zero except at  special positions
	called terminals, each of which only has a single outgoing arc forming a self-loop, we get a {\it stochastic terminal payoff} game,
	and when the self-loops have 0/1 payoffs, and every random position has only two outgoing arcs with probability $1/2$ each, we obtain the so-called
	\emph{simple stochastic games} (SSGs), introduced by Condon \cite{Con92,Con93} and considered in several papers \cite{GH08,Hal07}.
	In the latter games, the objective of \Max\  is to maximize the probability of reaching the terminal, while \Min\  wants to minimize this probability.
	Recently, it was shown that Gillette games with perfect information (and hence BWR-games by \cite{DGAA13}) are equivalent to SSGs under polynomial-time reductions \cite{AM09}.
	Thus, by results of Halman \cite{Hal07}, all these games can be solved
	in randomized strongly subexponential time $2^{O(\sqrt{n_d\log n_d})}$, where $n_d=|V_B|+|V_W|$ is the number of \emph{deterministic} vertices. On the other hand, if the number of random positions is constant, there are polynomial time algorithms for SSGs. Gimbert and Horn gave an $O(|V_R|! |V||E| + |p|)$ algorithm, where $|p|$ is the
	maximum bit-length of a transition probability. Chatterjee et al. \cite{CAH09} pointed out that a variant of strategy iteration can be
	implemented to solve SSGs in time $4^{|V_R|}|V_R|^{O(1)}|V|^{O(1)}$. 
	Dai and Ge \cite{DG09} gave
	a randomized algorithm with expected running time
	$\sqrt{|V_R|!}|V|^{O(1)}$.
	Ibsen-Jensen and Miltersen \cite{JM12} improved these bounds by showing that a variant of value iteration solves SSGs in time in $O(|V_R|2^{|V_R|}(|V_R|\log |V_R| +
	|V|))$. 
	For BW-games several pseudo-polynomial and subexponential algorithms are known 
	\cite{GKK88,KL93,ZP96,Pis99,BV01,BV01a,BSV04,BV05,BV07,Hal07,Vor08};
	see also \cite{JPZ06} for parity games.
	Besides their many applications (see e.g. \cite{Lit96,J2000}), all these games are of interest to Complexity Theory: Karzanov and Lebedev \cite{KL93} (see also \cite{ZP96}) proved that
	the decision problem ``whether the value of a BW-game is positive" is in the intersection of NP and co-NP. Yet, no polynomial algorithm is known for these games, see e.g., the survey by Vorobyov \cite{Vor08}. A similar complexity claim can be shown to hold for SSGs and BWR-games, see \cite{AM09,DGAA13}.
	
	\subsection{Main result}
	
	The problem of developing an efficient algorithm
	for stochastic games with perfect information was mentioned as an  open problem in the survey \cite{RF91}. While there are numerous pseudo-polynomial algorithms known for the BW-case, it is a challenging open question whether a pseudo-polynomial algorithm exists for BWR-games, as the existence of such an algorithm would imply also the polynomial solvability of this class of games \cite{AM09}. Our main result can be viewed as a partial solution of this problem, for the case when the number of random positions is fixed.
	
	For a BWR-game $\cG$ let us denote by $n=|V_W|+|V_B|+|V_R|$ the number of positions, by $k=|V_R|$ the number of random positions, and assume that all local rewards are integral with maximum absolute value $R$, and all transition probabilities are rational with common denominator $D$. The main result of this paper is as follows.
	
	\begin{theorem} \label{t-main}
		A BWR-game $\cG$ can be solved in $(nDk)^{O(k)}R\cdot \polylog R$ time.
	\end{theorem}
	
	As we shall see from our proof in Section~\ref{sec:non-ergodic}, one can replace $n^{O(k)}$ in the above theorem by $\poly(n)\nu(\cG)^{O(k)}$, where $\nu(\cG)$ is defined by a parametrized BW-game obtained from $\cG$ (see the precise definition in Section~\ref{sec:basic}). For stochastic terminal payoff games with $t$ terminals, it can be shown that $\nu(\cG)\le t+1$. Thus, Theorem \ref{t-main} extends the fixed-parameter tractability of simple stochastic games with respect to the number of random positions \cite{CAH09,DG09,GH08,JM12} and the pseudo-polynomial solvability of deterministic mean payoff games \cite{GKK88,Pis99,ZP96}.

	It is important to note that the above result requires a new technique to solve BWR-games. Approximating by discounted games cannot give the claimed complexity. The example in \cite{BEGM-ORL} shows that ones needs to choose $\beta$ exponentially (in $n$) close to 1 even if we have only a {\it single} random position.  
	
	Theorem~\ref{t-main} combined with the reduction in \cite{BEFGMM11} implies that we can obtain an $\eps$-saddle point (that is, a pair of stationary strategies that approximate the value within an error of $\eps$) for a BWR-game in time $\poly(n,D^k,\frac{1}{\eps})$.
	\subsection{Main ideas of the proof}
	\label{ss03}
	
	Our approach for proving Theorem~\ref{t-main} relies heavily on reducing the computation of uniformly optimal strategies for a general BWR-game to the case of {\it ergodic} BWR-games, i.e., those in which the (optimal) value does not depend on initial position, and showing that this special case can be solved in $(Dk)^{O(k)}R\cdot\poly(n,\log R)$ time. 
	
	\medskip
	
	Our algorithm for the ergodic case is based on potential transformations which change the local reward without changing the normal form of the game; see Section~\ref{ssPot}. 
	We would like to bring the local rewards to such a form that every locally optimal move is also globally optimal. Starting from zero potentials, the algorithm keeps selecting a subset of positions and reducing their potentials until either the locally optimal rewards (maximum for \Max, minimum for \Min, and average for Random) at different positions become sufficiently close to each other, or a proof of non-ergodicity is obtained in the form of a certain partition of the positions. In more details, the algorithm proceeds in phases. In one phase, we divide the range of current locally optimal rewards into four regions, and keep reducing the potentials of some of the positions such that no position can escape from the middle two regions. The phase ends when either the top or bottom region becomes empty, or a proof of non-ergodicity is found. Note that an algorithm for BW-games, also based on potential reductions, was suggested in \cite{GKK88}. However, as mentioned in \cite{GKK88}, this algorithm cannot be extended to BWR-games since in this algorithm the middle regions consist of exactly one level. Thus random positions with arcs going out of the middle level may have to escape from that level after the potential reduction. In our algorithm, we overcome this difficulty by relaxing the middle level to a set of levels.
	
	The upper bound on the running time consists of three technical parts. The first one is to show that if the number of iterations becomes too large, then there is a large enough potential gap to ensure non-ergodicity. In the second part, we show that the range of potentials can be kept sufficiently small throughout the algorithm, namely $\|x^*\|_{\infty}\leq n Rk(2D)^{k}$, and hence the range of the transformed rewards does not explode. The third part concerns the required accuracy. We show that it is enough to use an accuracy of
	\begin{equation}\label{accur}
	\eps< ((k+1)^2n^2(2D)^{2k+6})^{-1}
	\end{equation}
	in order to guarantee that the algorithm either finds the exact value or discovers non-ergodicity. As we shall see below, this accuracy is also enough to find an optimal strategy in the non-ergodic case. This contrasts with the fact that there is an example with $k=1$, in which the difference between two distinct values is exponentially small in $n$; see \cite{BEGM-ORL}. 
	We also show a lower bound of $D^{\Omega(k)}$ on the running time of the algorithm of 
	Theorem \ref{t-main} by constructing a series of games, with only random positions (that is, weighted Markov chains).
	
	\medskip
	
	For a non-ergodic game, the above algorithm can be used to find the classes of positions with the largest and smallest values, but cannot find other classes, since they interact via random positions.
	Note that a stochastic terminal game with $k$ random positions can be reduced, by guessing the order of values of the random positions, to $k!$ deterministic  reachability games, each of which can be solved in polynomial time~\cite{GH08}. In contrast, in the case of BWR-games, even if we know the actual values (not only the order) of all the random positions, it is not clear how to find optimal strategies realizing these values. Nevertheless, we show that if the values are known, then the problem can be reduced to the ergodic case. Then to utilize the algorithm for the ergodic case, we employ the idea of solving {\it parametrized games} 
	to reduce the search space for all values into a set of $
	k$-tuples of rational intervals of cardinality at most $\nu(\cG)^{O(k)}$. Using such a set of tuples we can iteratively replace random positions by self-loops on which the local reward is a guessed value. In  more details:
	
	\begin{enumerate}
		\item We iterate steps 2, 3 and 4 below over the random positions in the guessed order, keeping only the positions with highest rank (and hence having the same optimal value), and deleting all the other random positions. We iterate until no random positions remain, in which case we solve a BW-game.
		\item We consider the situation when all the kept random positions are replaced by a self-loop with local reward parameter $x$; we show that the value of any position in the resulting game defines an interval on the real line, as $x$ changes from $-\infty$ to $+\infty$. We identify a set of at most $\nu(\cG)+1$ maximal intervals, in each of which, the values of different positions as functions of $x$ are either constant or equal to $x$ in the entire interval.
		\item Since we do not know the real value of $x$, we guess among the identified intervals one that contains $x$; for the guessed interval, we provide optimal strategies for the  positions that have values above the lower bound of the interval, assuming our guess is correct.
		\item Each of our guesses above yields a pair of strategies that can be verified for optimality by solving two MDPs.
	\end{enumerate}
	Note that the number of guesses is bounded by $(k\nu(\cG))^{O(k)}$. 
	\begin{remark}
		It is interesting to note that in the BW-model the ergodic case is as difficult as the general one  \cite{GKK88} while in the BWR-model the non-ergodic case seems more difficult, see Section \ref{sec:non-ergodic}.   
	\end{remark}
	
	\section{Preliminaries}\label{s0}
	
	\subsection{Markov chains}
	\label{s-limit}
	Let $(G=(V,E),P)$ be a Markov chain, and let $C_1,\ldots,C_k\subseteq V$ be the vertex sets of the strongly connected components (classes) of
	$G$. For $i\neq j$, let us (standardly) write $C_i\prec C_j$
	if there is an arc $(v,u)\in E$
	such that $v\in C_i$ and $u\in C_j$. The components $C_i$, such that there is no $C_j$ with $C_i\prec C_j$ are called the \emph{absorbing} (or \emph{recurrent})
	classes, while the other components are called \emph{transient} or {\em non-recurrent}. 
	Let $J=\{i\mid C_i\mbox{ is absorbing}\}$, $A=\cup_{i\in J}C_i$, and $T=V\setminus A$.
	For $X,Y\subseteq V$, a matrix $H\subseteq\RR^{V\times V}$, a vector $h\subseteq\RR^V$, 
	we denote by $H[X;Y]$ the submatrix of $H$ induced by $X$ as rows and $Y$ as columns,
	and by $h[X]$ the subvector of $h$ induced by $X$. Let $I=I[V;V]$ be the $|V|\times|V|$ identity matrix, 
	$e=e[V]$ be the vector of all ones of dimension $|V|$. For simplicity,
	we drop the indices of $I[\cdot,\cdot]$ and $e[\cdot]$, when they are understood from the context.
	Then $P[C_i;C_j]=0$ if $C_j\prec C_i$, and hence in particular,
	$P[C_i;C_i]e=e$ for all $i\in J$, while $P[T,T]e$ has at least one component of value strictly less than $1$.
	
	The following are well-known facts about $P^i$ and the limiting distribution $p_w=e_wP^*,$ when the initial distribution is
	the $w$th unit vector $e_w$ of dimension $|V|$ (see, e.g., \cite{KS63}):
	\begin{itemize}
		\item[(L1)] $p_w[A]>0$ and $p_w[T]=0$;
		\item[(L2)] $\lim_{i\rightarrow\infty} P^i[V;T]=0$;
		\item[(L3)] $\rank(I-P[C_i;C_i])=|C_i|-1$ for all $i\in J$, $\rank(I-P[T;T])=|T|$, and $(I-P[T;T])^{-1}=\sum_{i=0}^\infty P[T;T]^i$;
		\item[(L4)] the absorption probabilities $y_i\in[0,1]^V$ into a class $C_i$, $i\in J$, are given by the unique solution of the linear system:
		$(I-P[T;T])y_i[T]=P[T;C_i]e$, $y_i[C_i]=e$ and $y_i[C_j]=0$ for $j \in J$ with $j\neq i$;
		\item[(L5)] the limiting distribution $p_w\in[0,1]^V$ is given by the unique solution of the linear system: $p_w[C_i](I-P[C_i;C_i])=0, ~p_w[C_i]e=y_i(w)$, for all $i\in J$, and $p_w[T]=0$.
	\end{itemize}
	
	\subsection{BWR-games, solvability and ergodicity}
	\label{sss01}
	
	A BWR-game $\cG = (G, p, r)$ is given by a digraph $G = (V, E)$, where $V=V_W\cup V_B\cup V_R$ is a partition of the vertices;
	$G$ may have loops and multiple arcs, but no terminal positions, i.e., positions of out-degree $0$; $p(v,u)$ are probabilities for $v\in V_R$, $(v,u)\in E$ satisfying $\sum_{u \; | \; (v, u) \in E} p(v, u) = 1$ for all $v \in V_R;$ and  $r:E\to \ZZ$ is a local reward function. For convenience we will also assume that
	$p(v, u) > 0$  whenever
	$(v, u) \in E$  and  $v \in V_R$, and
	set  $p(v, u) = 0$
	for  $(v, u) \not \in E$.
	
	
	Standardly, we define a strategy  $s_W\in S_W$
	(respectively,  $s_B\in S_B$) as a mapping
	that assigns a position  $u \in V$, such that $(v,u)\in E$, to each position
	$v \in V_W$  (respectively,  $v \in V_B$).
	A pair of strategies  $s = (s_W, s_B)$
	is called a {\em situation}.
	Given a BWR-game  $\cG = (G, p, r)$  and
	situation  $s = (s_B, s_W)$,  we obtain
	a weighted Markov chain $\cG_s = (P_s, r)$ with transition matrix $P_s$ in the obvious way:
	\begin{eqnarray*}
		\label{extend}
		p_s(v,u) =\left\{
		\begin{array}{ll}
			1&\mbox{  if } (v \in V_W \mbox{ and } u = s_W(v)) \mbox{ or } (v \in V_B \mbox{  and }u = s_B(v));\nonumber\\
			0&\mbox{ if } (v \in V_W\mbox{   and  }u \neq s_W(v))\mbox{   or } (v \in V_B\mbox{   and  }u \neq s_B(v));\\
			p(v,u)&\mbox{ if } v\in V_R.
		\end{array}
		\right.
	\end{eqnarray*}
	In the obtained Markov chain
	$\cG_s = (P_s, r)$, we define
	the limiting (mean) effective payoff $\mu_{\cG_s}(v)=\mu_s(v)$ as
	\begin{equation}\label{value}
	\mu_s(v)=\sum_{w\in V}p^*_s(v,w)\sum_{u}p_s(w,u)r(w,u),
	\end{equation}
	where $p^*_s(v,w)$ is the limit probability in $\cG_s$ to be at position $w$ when the initial position is $v$ (see Section~\ref{s-limit} for more details). 
	It is known \cite{Gil57,LL69} that
	every such game has a pair of (uniformly optimal) pure stationary strategies $(s^*_W,s^*_B)$  such that for any other pair of stationary strategies $(s_W,s_B)$ and for every initial position $v$, the following hold:
	$$
	\mu_{(s_W,s_B^*)}(v)\le\mu_{(s_W^*,s_B^*)}(v)\le\mu_{(s_W^*,s_B)}(v).
	$$ 
	The quantity $\mu_{(s_W^*,s_B^*)}(v)$ is called the {\it value of the game} starting from position $v$, and will be denoted by $\mu_{\cG}(v)$, or simply by $\mu(v)$ if the game is clear from the context.
	The value $\mu_{\cG}(v)$ may depend on  $v$.
	The BWR-game  $\cG = (G, p, r)$   is called
	{\em ergodic} if the value $\mu(v)$ is the same
	for all initial positions  $v\in V$.

	\subsection{Potential transformations and canonical forms}
	\label{ssPot}\label{ssCanon}
	Given a BWR-game  $\cG = (G, p, r)$, let us introduce
	a mapping  $x : V \rightarrow \RR$, whose
	values  $x(v)$  will be called {\em potentials}, and
	define the transformed reward function $r_x:E\to\RR$  as:
	\begin{equation}
	\label{eqpot}
	r_x(v,u) = r(v,u) + x(v) - x(u),
	\;\; \mbox{where} \;\;
	(v,u) \in E.
	\end{equation}
	
	\smallskip
	
	It is not difficult to verify that the two normal forms of the obtained game  $\cG^x$
	and the original game  $\cG$, are the same, and hence the games $\cG^x$ and $\cG$ are equivalent (see \cite{DGAA13}).
	In particular, their optimal
	(pure stationary) strategies coincide, and
	their value functions also coincide: $\mu_{\cG^x} = \mu_{\cG}$.

	It is known that for BW-games there exists a potential transformation such that, in the obtained game the locally optimal strategies are globally optimal,
	and hence, the value and optimal strategies become obvious \cite{GKK88}. This result was extended for the more general class of BWR-games in \cite{DGAA13}: in the transformed game, the equilibrium value $\mu_{\cG}(v)=\mu_{\cG^x}(v)$ is given simply by the maximum local reward for $v\in V_W$, the minimum local reward for $v\in V_B$, and the average local reward for $v\in V_R$. In this case we say that the transformed game is in \emph{canonical} form. To define this more formally, let us use the following notation throughout this section: Given functions $f:E\to\RR$ and $g:V\to\RR$, we define the functions $\bar{M}[f],\bar{M}[g]:V\to\RR$.
	
	\begin{equation*}
	\bar{M}[f](v) = \left\{\begin{array}{lll}
	\max_{u \mid (v, u) \in E}f(v,u), &\text{for } v \in V_W, \\
	\min_{u \mid (v, u) \in E}f(v,u), &\text{for } v \in V_B, \\
	\sum_{u \mid (v, u) \in E}   p(v,u)\, f(v,u), &\text{for } v \in V_R.
	\end{array}\right.
	\end{equation*}
	
	\begin{equation*}
	\bar{M}[g](v) = \left\{\begin{array}{lll}
	\max_{u \mid (v, u) \in E}g(u), &\text{for } v \in V_W, \\
	\min_{u \mid (v, u) \in E}g(u), &\text{for } v \in V_B, \\
	\sum_{u \mid (v, u) \in E}   p(v,u)\, g(u), &\text{for } v \in V_R.
	\end{array}\right.
	\end{equation*}
	
	We say that a BWR-game $\cG$ is in canonical form if there exist vectors $\mu,x\in\RR^V$ such that
	\begin{itemize}
		\item[(C1)]   $\mu= \bar M[\mu]=\bar M[r_x]$ and,
		\item[(C2)] for every  $v \in V_W \cup V_B$,
		every move  $(v,u) \in E$
		such that  $\mu(v) =  r_x(v,u)$ must also have $\mu(v)=\mu(u)$, or
		in other words, every locally optimal move $(v,u)$
		is globally optimal.
	\end{itemize}
	
	
	Canonical forms were defined for BW-games in \cite{GKK88}, and extended to
	BWR-games and other more classes of stochastic games in \cite{DGAA13}.
	It was shown in \cite{GKK88} that there always exists a potential transformation $x$ such that the optimal local rewards $\bar M[r_x](v)$ in a BW-game are equal to the game's value $\mu(v)$ at each vertex $v$. This result was extended
	in \cite{DGAA13} to the BWR-case. 
	
	\begin{theorem}[\cite{DGAA13}]
		\label{t2}
		For each BWR-game  $\cG$, there
		is a potential transformation
		\raf{eqpot} bringing  $\cG$  to
		canonical form. Furthermore, in a game in canonical form we have $\mu_{\cG} = \bar M[r]$.
	\end{theorem}
	
	%
	%
	%
	
	In this paper we will provide an algorithm for finding such a potential transformation in the ergodic case.
	
	\begin{proposition}
		\label{p1-}
		If there exists a constant $m$ such that $\bar M[r]=m$ for all $v \in V$, 
		then
		(i) every locally optimal strategy is optimal and
		(ii) the game is ergodic:
		$m=\mu(v)$  is its value for every initial position  $v\in V$.
	\end{proposition}
	\proof
	Indeed, if \Max\ (\Min) applies a locally optimal strategy
	then after every own move (s)he will get
	(pay)  $m$,  while for each move of the opponent
	the local reward will be at least (at most)  $m$, and
	finally, for each random position the expected
	local reward is  $m$.
	Thus, every locally optimal strategy of a player is optimal.
	Furthermore, if both players choose their optimal strategies
	then the expected local reward
	$\EE[r(e_i)]$  equals  $m$  for every step  $i$.
	Hence, the  value of the game
	$\lim_{T\rightarrow\infty}\frac{1}{T+1}\sum_{i=0}^T b_i$ equals  $m$.
	\qed
	
	\subsection
	{Sufficient conditions for ergodicity of BWR-games}
	\label{sErgo}
	
	A digraph   $G = (V=V_W\cup V_B\cup V_R, E)$
	is called
	{\em ergodic} if all BWR-games  ${\bf \cG} = (G,p, r)$
	on $G$ are ergodic. We will give a simple characterization
	of ergodic digraphs, which,
	obviously, provides a sufficient condition
	for ergodicity of BWR-games.
	For BW-games
	(that is, in case $R = \emptyset$)  such
	characterization was given in \cite{GL89}.
	
	\smallskip
	
	In addition to partition
	$V = V_W \cup V_B \cup V_R$, let us
	consider another partition
	$\Pi: V = V^+ \cup V^- \cup V^0$
	with the following properties:
	
	\begin{description}
		\item[(i)]  Sets  $V^+$  and  $V^-$  are not empty
		(while  $V^0$  might be empty).
		
		\item[(ii)] There is no arc  $(v, u) \in E$  such that either
		$v \in (V_W \cup V_R) \cap V^-$  and  $u \not \in V^-$,
		or $v \in (V_B \cup V_R) \cap V^+$  and  $u \not \in V^+$.
		In other words,
		\Max\ cannot leave  $V^-$,
		\Min\ cannot leave  $V^+$, and
		there are no random moves leaving $V^+$ or $V^-$.
		
		\item[(iii)] For each  $v \in V_W \cap V^+$
		(respectively,  $v \in V_B \cap V^-$)  there is
		a move  $(v, u) \in E$  such that  $u \in V^+$
		(respectively,  $u \in V^-$).
		In other words, \Max\ 
		(\Min) cannot be forced to leave $V^+$
		(respectively,  $V^-$).
	\end{description}
	
	In particular, the properties above imply that the induced
	subgraphs  $G[V^+]$  and  $G[V^-]$
	have no terminal vertices. 
	
	\smallskip
	
	A partition  $\Pi: V = V^+ \cup V^- \cup V^0$
	satisfying  (i), (ii), and (iii)
	will be called a {\em contra-ergodic} partition for
	digraph $G = (V_W\cup V_B\cup V_R, E)$.
	
	\begin{theorem}
		\label{tErgo}
		A digraph  $G$  is ergodic iff 
		it has no contra-ergodic partition.
	\end{theorem}
	\proof
	{\em ``Only if part''}.
	Let  $\Pi: V = V^+ \cup V^- \cup V^0$
	be a contra-ergodic partition of   $G$.
	Let us assign arbitrary
	positive probabilities
	to random moves  such that
	$\sum_{u \; | \; (v,u) \in E} p(v, u) = 1$
	for all  $v \in V_R$.
	We still have to assign a local reward
	$r(v, u)$  to each move  $(v, u) \in E$.
	Let us define
	$r(v, u) = 1$  whenever  $v, u \in V^+$,
	$r(v, u) = -1$ whenever  $v, u \in V^-$, and
	$r(v, u) = 0$  otherwise.
	Clearly, if the initial position is in  $V^+$
	(respectively,  in  $V^-$)  then
	the value of the obtained game is  $1$
	(respectively,  $-1$).
	Hence, the corresponding game
	is not ergodic.
	
	\medskip
	
	{\em ``If part''}.
	Given a non-ergodic BWR-game ${\bf \cG} = (G, P, r)$, the value function  $\mu_\cG$ is not constant.
	Let  $\mu_W$  and  $\mu_B$  denote
	the maximum and minimum values, respectively.
	Then, let us set
	$V^+ = \{v \in V \; | \; \mu(v) = \mu_W\}$,
	$V^- = \{v \in V \; | \; \mu(v) = \mu_B\}$, and
	$V^0 = V \setminus (V^+ \cup V^-)$.
	It is not difficult to verify that
	the obtained partition
	$\Pi : V = V^+ \cup V^- \cup V^0$  is contra-ergodic.
	\qed

	The ``only if part'' can be strengthened as follows.
	
	\medskip
	
	A \emph{contra-ergodic decomposition} of $\cG$ is a contra-ergodic partition
	$\Pi : V = V^+ \cup V^- \cup V^0$ such that
	$\bar M[r](v) > \bar M[r](u)$  for every  $v \in V^+$  and  $u \in V^-$.
	
	\begin{proposition}
		\label{pErgo}
		Given a BWR-game  $\cG$  whose
		graph has a contra-ergodic partition,
		if  $\bar M[r](v) > \bar M[r](u)$
		for every  $v \in V^+, u \in V^-$
		then  $\mu(v) > \mu(u)$
		for every  $v \in V^+, u \in V^-$. In particular, $\cG$ is not ergodic. 
	\end{proposition}
	\proof
	Let us choose a number  $\mu$  such that
	$\bar M[r](v) > \mu > \bar M[r](u)$
	for every  $v \in V^+$  and  $u \in V^-$;
	it exists, because set $V$ of positions is finite.
	Obviously, properties (i), (ii), and (iii) imply that
	\Max\ (\Min) can guarantee
	more  (less) that $\mu$
	for every initial position  $v \in V^+$
	(respectively, $v \in V^-$).
	Hence, $\mu(v) > \mu > \mu(u)$
	for every  $v \in V^+$  and  $u \in V^-$.
	\qed

	For example, no contra-ergodic partition can exist if
	$G = (V_W\cup V_B\cup V_R, E)$  is a complete tripartite
	digraph.
	
	
	\section
	{Ergodic BWR-games}
	\label{ErgodPump}
	
	\subsection{Description of the pumping algorithm}
	\label{ErgodPump1}
	
	Given a BWR-game  $\cG = (G, p, r)$, let us
	compute  $m(v)=\bar M[r](v)$  for all  $v \in V$. 
	The algorithm proceeds in phases. In each phase we iteratively change the potentials such that either the range of $m$ is reduced by a factor of 3/4, or we discover a contra-ergodic partition. 
	The starting local reward function in the next  phase is obtained by transforming the initial rewards by the potential function obtained in the current phase. Once the range becomes small enough we stop repeating these phases, which allows us to conclude that the game is ergodic.
	
	Now we describe informally the steps in one phase. The general step of a phase, called {\it pumping}, consists of reducing all potentials of  the positions with $m$-values in the upper half of the range by the same constant $\delta$; we say in this case that those positions are {\it pumped}. 
	It will be not difficult  to show that the $m$-values of all positions in this upper half can only decrease and by at most $\gd$, while the $m$-values of all other positions can only increase and also by at most $\gd$. In each step we will choose $\gd$ as the largest constant such that no $m$-value leaves the middle half (that is, second and third quarter) of the range. 
	It can happen that $\gd=+\infty$; in this case, we can construct a contra-ergodic decomposition proving that the game is non-ergodic. 
	
	Next, we give a more formal description of phase $h=0,1,\ldots$ Let us denote by $x^{h}$ the potential at the end of phase $h-1$, where we set $x^0=0$. 
	Given a function $f:V\to\RR$, let us denote by $f^+$ and $f^-$ its maximum and minimum values respectively. Throughout, we will denote
	by $[m]=[m^-,m^+]$ and $[r]=[r^-,r^+]$ the range of functions $m$ and $r$, respectively, and let $M=m^+-m^-$ and $R=r^+-r^-$. 
	Given a potential function $x:V\to\RR$, we will denote by $m_x$, $M_x$, etc., the functions $m$, $M$, etc., with $r$ replaced by the transformed reward $r_x$ in the above definitions. 
	Given a subset  $I \subseteq [m]$,  let
	$V_x(I) = \{v \in V \; | \; m_x(v) \in I\} \subseteq V$.
	In the following algorithm, set  $I$
	will always be a closed or semi-closed
	interval within  $[m]$ (and not within $[m_x]$).
	For convenience, we will write $(\cdot)_{x^h}$ as $(\cdot)_h$, 
	where $(\cdot)$ could be $m$, $r$, $r^+$, etc (e.g., $m_h^-=m_{x^h}^-$, $m_h^+=m_{x^h}^+$).

	Let $m^-_h=t_0<t_1<t_2<t_3<t_4=m^+_h$ be thresholds defined by
	%
	\begin{equation}\label{thresholds}
	t_i=m^-_h+\frac{i}{4}M_h, ~~i=0,1,2,3,4,~~\mbox{ where $M_h=m_h^+-m_h^-$.}
	\end{equation}

	
	Let us introduce $\hat x(v)=x(v)-\delta$ for $v\in V_x[t_2,t_4]$ and $\hat x(v)=x(v)$ for $v\in V_x[t_0,t_2)$. Let us then introduce the notation $m_x^{\gd}$ for $m_{\hat x}$. This notation may look complicated but is necessary since $m_{\hat x}$ depends on $x$ and $\delta$ in a complicated way.
	
	It is clear that  $\gd$
	can be computed in linear time, and we have $m_x^{\gd}(v)\ge t_1$ for all $v\in V_x[t_2,t_4]$ and $m_x^{\gd}(v)\le t_3$ for all $v\in V_x[t_0,t_2)$, where $m_x^{\gd}(v)$ is the new value of $m_x(v)$ after all potentials in $V_x[t_2, t_4]$ have been reduced by $\delta$. The value of $m_x^{\gd}$ is given by the following formula:
	
	{\small
		\begin{eqnarray*}\label{updates1}
			\max\left\{\max_{(v,u)\in E,\atop u\in V_x[t_2,t_4]}\{r_x(v,u)\}, \max_{(v,u) \in E,\atop u\in V_x[t_0,t_2)}\{r_x(v,u)\}-\gd\right\} &\text{for }& v \in V_W\cap V_x[t_2,t_4],\nonumber\\
			\min\left\{\min_{(v,u)\in E,\atop u\in V_x[t_2,t_4]}\{r_x(v,u)\}, \min_{(v,u) \in E,\atop u\in V_x[t_0,t_2)}\{r_x(v,u)\}-\gd\right\} &\text{for }& v \in V_B\cap V_x[t_2,t_4],\nonumber\\
			\sum_{(v,u)\in E,\atop u\in V}p(v,u)r_x(v,u)-\gd\sum_{(v,u)\in E,\atop u\in V_x[t_0,t_2)}p(v,u) &\text{for }& v \in V_R\cap V_x[t_2,t_4],\nonumber
		\end{eqnarray*}
		
		\begin{eqnarray*}\label{updates2}
			\max\left\{\max_{(v,u)\in E,\atop u\in V_x[t_2,t_4]}\{r_x(v,u)\}+\gd, \max_{(v,u) \in E,\atop u\in V_x[t_0,t_2)}\{r_x(v,u)\}\right\} &\text{for }& v \in V_W\cap V_x[t_0,t_2),\nonumber\\
			\min\left\{\min_{(v,u)\in E,\atop u\in V_x[t_2,t_4]}\{r_x(v,u)\}+\gd, \min_{(v,u) \in E,\atop u\in V_x[t_0,t_2)}\{r_x(v,u)\}\right\} &\text{for }& v \in V_B\cap V_x[t_0,t_2),\nonumber\\
			\sum_{(v,u)\in E,\atop u\in V}p(v,u)r_x(v,u)+\gd\sum_{(v,u)\in E,\atop u\in V_x[t_2,t_4]}p(v,u) &\text{for }& v \in V_R\cap V_x[t_0,t_2).\nonumber\\
		\end{eqnarray*}
	}
	Note that $|m_x^{\gd}(v)-m_x(v)|\le\delta$. 
	It is also important to note that  $\gd \geq M_h/4$.
	Indeed, the $m$-values of positions from  $V_x[t_2, t_4]$ cannot increase,
	while  those of positions from  $V_x[t_0, t_2)$ cannot decrease.
	Each of them would have to traverse a distance of at least  $M_h/4$
	before it can reach the border of the interval
	$V_x[t_1, t_3]$.
	Moreover, if after some iteration one of the sets
	$V_x[t_0, t_1)$  or  $V_x(t_3, t_4]$  becomes empty
	then the range of  $m_x$  is reduced at least by  $25\%$.
	
	
	Procedure PUMP$(\cG,\eps)$ below tries to reduce any BWR-game $\cG$ by a potential transformation $x$ into one in which $M_x\le\eps$.
	Two subroutines are used in the procedure. REDUCE-POTENTIALS$(\cG,x)$ replaces the current potential $x$ with another potential with a sufficiently small norm; see Lemma \ref{pot-red} in Subsection \ref{sec:pot-red}. This reduction is needed since without it the potentials and, hence, the transformed local rewards too, may grow exponentially; see the analysis of $R_h$ and $N_h$ in the proof of Lemma \ref{l-run}. The second routine FIND-PARTITION$(\cG,x)$ uses the current potential vector $x$ to construct a contra-ergodic decomposition of $\cG$ 
	(cf. line \ref{find-part} of the algorithm below), see Subsection \ref{sec:find-part} for the details.
	
	Note the algorithm can terminate in 3 ways:  
	\begin{enumerate}
		\item $\delta=+\infty$ in some iteration. In this case, $(V^+=V_x[t_0,t_2)$, $V^-=V_x[t_2,t_4]$, $V^0=\emptyset)$ is a contra-ergodic partition as can be checked from the definition of $m^\gd_x$. 
		\item The number of pumping iterations performed in some phase $h$ reaches 
		\begin{equation}\label{N}
		N_h=\left\lfloor\frac{8n^2R_hD^k}{M_{h}}\right\rfloor+1,
		\end{equation}
		where $R_h=r_h^+-r_h^-$, and yet the range of $m_h$ is not reduced.
		In this case, we can find a contra-ergodic decomposition by the second subroutine FIND-PARTITION$(\cG,x)$; see Lemma \ref{l-gap}. 
		\item $M_{x}\leq \eps$ and $\eps$ satisfies \raf{accur}. In this case we can prove that a pair of locally optimal strategies with respect to $r_x$ is optimal in the game $\cG$; we show this in Section~\ref{ss21}. 
	\end{enumerate}
	
	A simple example illustrating how the pumping algorithm works is given below.
	
	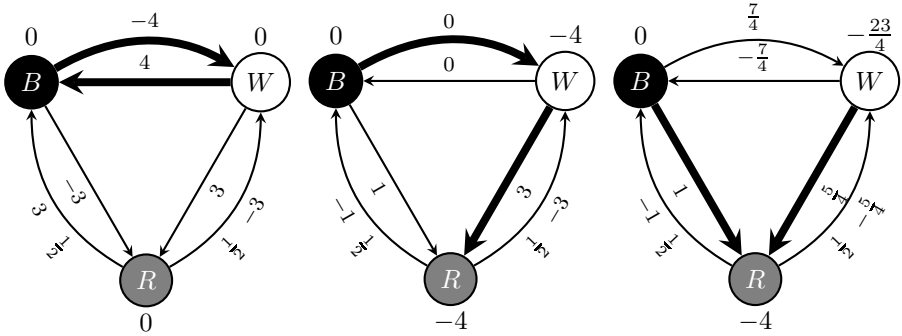
\begin{figure}[h]
		\begin{center}
			\begin{tikzpicture}[>=stealth,scale=0.6]
			\node at (1in,0) [shape=circle,minimum size=10pt,draw=black,thick] (W) {$W$};
			\node at (-1in,0) [shape=circle,minimum size=10pt,draw=black,thick,fill=black,text=white] (B) {$B$};
			\node at (0,-1.73in) [shape=circle,minimum size=10pt,draw=black,thick,fill=gray,text=white] (R) {$R$};
			\node at (1in,0.4in) {$0$};
			\node at (-1in,0.4in) {$0$};
			\node at (0,-2.1in) {$0$};
			\path[->,line width=3pt]
			(W) edge node[black,sloped,above,midway] {\footnotesize $4$} (B)
			(B) edge[bend left] node[black,sloped,above,midway] {\footnotesize $-4$} (W);
			\path[->,thick]  
			(B) edge node[black,sloped,below,midway] {\footnotesize $-3$} (R)
			(R) edge[bend left] node[black,sloped,below,pos=0.3] {\footnotesize $\frac12$} node[black,sloped,below,midway] {\footnotesize $3$} (B)
			(W) edge node[black,sloped,below,midway] {\footnotesize $3$} (R)
			(R) edge[bend right] node[black,sloped,below,pos=0.3] {\footnotesize $\frac12$} node[black,sloped,below,midway] {\footnotesize $-3$} (W);
			\end{tikzpicture}$~$
			\begin{tikzpicture}[>=stealth,scale=0.6]
			\node at (1in,0) [shape=circle,minimum size=10pt,draw=black,thick] (W) {$W$};
			\node at (-1in,0) [shape=circle,minimum size=10pt,draw=black,thick,fill=black,text=white] (B) {$B$};
			\node at (0,-1.73in) [shape=circle,minimum size=10pt,draw=black,thick,fill=gray,text=white] (R) {$R$};
			\node at (1in,0.4in) {$-4$};
			\node at (-1in,0.4in) {$0$};
			\node at (0,-2.1in) {$-4$};
			\path[->,line width=3pt]
			(B) edge[bend left] node[black,sloped,above,midway] {\footnotesize $0$} (W)
			(W) edge node[black,sloped,below,midway] {\footnotesize $3$} (R);
			\path[->,thick]
			(W) edge node[black,sloped,above,midway] {\footnotesize $0$} (B)
			(B) edge node[black,sloped,below,midway] {\footnotesize $1$} (R)
			(R) edge[bend left] node[black,sloped,below,pos=0.3] {\footnotesize $\frac12$} node[black,sloped,below,midway] {\footnotesize $-1$} (B)
			(R) edge[bend right] node[black,sloped,below,pos=0.3] {\footnotesize $\frac12$} node[black,sloped,below,midway] {\footnotesize $-3$} (W);
			\end{tikzpicture}$~$
			\begin{tikzpicture}[>=stealth,scale=0.6]
			\node at (1in,0) [shape=circle,minimum size=10pt,draw=black,thick] (W) {$W$};
			\node at (-1in,0) [shape=circle,minimum size=10pt,draw=black,thick,fill=black,text=white] (B) {$B$};
			\node at (0,-1.73in) [shape=circle,minimum size=10pt,draw=black,thick,fill=gray,text=white] (R) {$R$};
			\node at (1in,0.4in) {$-\frac{23}{4}$};
			\node at (-1in,0.4in) {$0$};
			\node at (0,-2.1in) {$-4$};
			\path[->,line width=3pt]
			(W) edge node[black,sloped,below,midway] {\footnotesize $\frac54$} (R)
			(B) edge node[black,sloped,below,midway] {\footnotesize $1$} (R);
			\path[->,thick]
			(W) edge node[black,sloped,above,midway] {\footnotesize $-\frac74$} (B)
			(B) edge[bend left] node[black,sloped,above,midway] {\footnotesize $\frac74$} (W)
			(R) edge[bend left] node[black,sloped,below,pos=0.3] {\footnotesize $\frac12$} node[black,sloped,below,midway] {\footnotesize $-1$} (B)
			(R) edge[bend right] node[black,sloped,below,pos=0.3] {\footnotesize $\frac12$} node[black,sloped,below,midway] {\footnotesize $-\frac54$} (W);
			\end{tikzpicture}
			
		\end{center}
		\caption{\label{fig000}An example with showing two iterations with the pumping algorithm.}
	\end{figure}
	
	Let us consider the example shown in the left of Figure \ref{fig000}. 
	Positions W and B are controlled by the maximizer and minimizer, respectively. Position R is a random position with probability $\frac12$ on both outgoing arcs. 
	
	We assume that the potentials, shown next to the vertices in Figure \ref{fig000},  initially are $x_W=x_B=x_R=0$. 
	
	In the first $h=0$ iteration of the pumping algorithm we have $m_x(W)=4$, $m_x(B)=-4$ and $m_x(R)=0$. The extremal arc from the black and white vertices are shown as thick lines in Figure \ref{fig000}. 
	Thus we have $M_0=8$ and hence $t_0=-4$, $t_1=-2$, $t_2=0$, $t_3=2$ and $t_4=4$. We get $V_x[t_2,t_4]=\{W,R\}$ and as a maximal pumping move $\delta=4$. 
	Thus we update $x_W=-4$, $x_R=-4$ and leave $x_B=0$ unchanged. The middle part in Figure \ref{fig000} shows the resulting graph, with the updated local rewards. 
	Since $V_x[t_0,t_1)=\emptyset$, we need to recompute the range. Therefore, in the next $h=1$ step we get 
	$m_x(W)=3$, $m_x(B)=0$ and $m_x(R)=-2$, yielding $M_1=5$, and $t_0=-2$, $t_1=-\frac34$, $t_2=\frac12$, $t_3=\frac74$, and $t_4=3$. Thus $V_x[t_2,t_4]=\{W\}$, and the maximal pumping distance is $\delta=\frac74$. Consequently, we update the potential $x_W=-\frac{23}{4}$, and leave the other two potentials unchanged. The right part in Figure \ref{fig000} shows the resulting graph. Since now $V_x(t_3,t_4]=\emptyset$, we will again recompute the range parameters. We get $m_x(W)=\frac54$, $m_x(B)=1$, and $m_x(R)=-\frac98$. Hence $M_2=\frac{19}{8}$. We leave it to the reader to follow the pumping algorithm on this small illustrative example.

	\medskip
	
	To describe the main idea of the analysis, we will first argue in Section \ref{ErgodPump2} that the algorithm terminates in finite time if the considered BWR-game is ergodic, even if we set $N_0=+\infty$. In the following section, this argument will be made quantitative with the precise bound on the running time.
	In Section \ref{Exp-ex}, we will
	show that our bound on the running time is tight since we can construct a weighted Markov chain (that is, an R-game) providing an exponential lower bound in $k$.

	\begin{algorithm}[t]
		\caption{PUMP$(\cG,\eps)$}
		\label{algo1}
		\begin{algorithmic}[1]
			\REQUIRE A BWR-game $\cG=(G=(V,E),P,r)$ and a desired accuracy $\eps$
			\ENSURE Either a potential $x:V\to\RR$ s.t. $|m_x(v)-m_x(u)|\leq\eps$ for all $u,v\in V$, or
			a contra-ergodic decomposition
			\STATE let $h:=0$; $x=x^h:=0$; $i:=1$
			\STATE let $t_0,t_1,\ldots,t_4$, and $N_h$ be as given by \raf{thresholds} and \raf{N}
			\WHILE{$i<N_h$}
			\IF {$M_{x}\leq \eps$}
			\RETURN $x$
			\ENDIF
			\STATE $\gd:=\max\{\gd'|~m_x^{\gd'}(v)\ge t_1\mbox{ for all }v\in V_{x}[t_2,t_4]\mbox{ and }m_x^{\gd'}(v)\le t_3\mbox{ for all }v\in V_{x}[t_0,t_2)\}$
			\IF {$\gd=\infty$}
			\label{cep1}     \RETURN the contra-ergodic partition $(V^+=V_x[t_0,t_2)$, $V^-=V_x[t_2,t_4]$, $V^0=\emptyset)$
			\ENDIF
			\STATE $x(v):=x(v)-\gd$ for all $v\in V_x[t_2,t_4]$
			\IF {$V_{x}[t_0,t_1)=\emptyset$ or $V_{x}(t_3,t_4]=\emptyset$}
			\label{reduced?}
			\STATE $x^{h+1}:=x$:=REDUCE-POTENTIALS$(\cG,x)$; $h=h+1$; $i:=1$
			\STATE recompute the thresholds $t_0,t_1,\ldots,t_4$ and $N_h$ using \raf{thresholds} and \raf{N}
			\ELSE
			\STATE $i:=i+1$; 
			\ENDIF
			\ENDWHILE
			\STATE $(V^+, V^-, V^0)$:=FIND-PARTITION$(\cG,x)$
			\label{find-part}
			\RETURN the contra-ergodic partition $(V^+, V^-, V^0)$
		\end{algorithmic}
	\end{algorithm}

	\subsection{Proof of finiteness for the ergodic case}
	\label{ErgodPump2}
	To simplify notation, we can assume without loss of generality (by shifting and scaling the local rewards) that the range of $m$ is $[0,1]$, that is, $t_i=\frac{i}{4}$, for $i=0,\ldots,4,$ and that the initial potential $x^0=0$.
	Suppose indirectly that phase $0$ of the algorithm does not terminate, that is, $V[0,\frac{1}{4})$ and $V(\frac{3}{4},1]$ never become empty and the $m$-range never gets smaller than $\frac{1}{2}$.
	
	Consider the infinite sequence of iterations and
	denote by  $V^- \subseteq V$
	(respectively, by  $V^+ \subseteq V$)  the set of vertices
	that were pumped just finitely many times
	(respectively, always but finitely many times);
	in other words, $m_{x}(v) \in [\frac{1}{2}, 1]$  if  $v \in V^+$
	(respectively,  $m_{x}(v) \in [0, \frac{1}{2})$  if  $v \in V^-$)
	for all but finitely many  iterations.
	By the above assumption, we have
	\begin{eqnarray}\label{ee1--}
	\emptyset\ne V[0,\frac{1}{4})\subseteq V^-\subseteq V[0,\frac{1}{2}),\\
	\label{ee2--}
	\emptyset\ne V(\frac{3}{4},1]\subseteq V^+\subseteq V[\frac{1}{2},1].
	\end{eqnarray}
	\begin{proposition}
		The partition  $\Pi : V = V^+ \cup V^- \cup V^0$, where
		$V^0 = V \setminus (V^+ \cup V^-)$, is a contra-ergodic decomposition.
	\end{proposition}
	\proof
	First, let us check properties (i), (ii), (iii)  of Section \ref{sErgo}. First, (i) follows by \raf{ee1--} and \raf{ee2--}. Let us observe that the transformed local reward on any arc leaving $V^+$ (resp., $V^-$) are $=+\infty$ (resp.,  is $-\infty$). This implies (ii) and (iii). 
	
	Finally, it follows from \raf{ee1--} and \raf{ee2--} also that 
	$m_{x}(v) >    \frac{1}{2}$  for all  $v \in V^+$, while
	$m_{x}(v) \leq \frac{1}{2}$  for all  $v \in V^-$. Hence the claim follows by Proposition \ref{pErgo}.  
	\qed
	
	In other words, our algorithm is finite for the ergodic BWR-games.
	Below we shall give an upper bound for the number of iterations a vertex can ``oscillate" in $[0,\frac{1}{4})$ or $(\frac{3}{4},1]$
	before it finally enters $[\frac{1}{4}, \frac{3}{4}]$ (to stay there forever).
	
	\begin{remark}
		For general stochastic games (with imperfect information), the value might not exist \cite{Gil57}.
		Nevertheless, the pumping algorithm can be extended \cite{BEGM14}. 
		This algorithm either (i) certifies that the game is $24\eps$-ergodic, 
		that is, it finds a potential transformation $x$ such that all local values $m_x(v)$ are within an interval of length $24\eps$, or 
		(ii) presents a contra-ergodic partition similar to the one described in Section~\ref{sErgo}; in particular, we obtain two positions $u$ and $v$ such that $|\mu(u)-\mu(v)|>\eps$. 
	\end{remark}
	
	
	\subsection{Finding a contra-ergodic decomposition: FIND-PARTITION$(\cG,x)$}\label{sec:find-part}
	We assume throughout this section that we are inside phase $h$ of the algorithm, which started with a potential $x^h$, and we proceed to step \ref{find-part}. 
	For simplicity, we assume that the phase starts with local reward function $r=r_h$ and hence\footnote{in particular, note that $r_x(v,u)$ and $m_x(v)$ are used, for simplicity of notation, to actually mean $r_{x+x^h}(v,u)$ and $m_{x+x^h}(v)$, respectively.} 
	$x^h=0$.
	Given a potential vector $x$, we use the following notation \cite{GKK88}:
	$$
	\EXT_x=\{(v,u)\in E\mid v\in V_B\cup V_W\mbox{ and }r_x(v,u)=m_x(v)\},
	$$
	and recall that $x^-=\min\{x(v)\mid v\in V\}.$
	Let $t_l \le 0$ be the largest value satisfying the following conditions:
	\begin{itemize}
		\item[(i)] there are no arcs $(v,u)\in E$ with $v\in V_W\cup V_R$, $x(v)\geq t_l$ and $x(u)<t_l$;
		\item[(ii)] there are no arcs $(v,u)\in\EXT_x$ with $v\in V_B$, $x(v)\geq t_l$ and $x(u)<t_l$.
	\end{itemize}
	Let $X=\{v\in V\mid x(v)\ge t_l\}$. In words, $X$ is 
	the set of positions with potential as close to $0$ as possible, such that 
	no white or random position in $X$ has an arc crossing to $V\setminus X$, and no black position has an extremal arc crossing to $V\setminus X$.
	Similarly, define $t_u\ge x^-$ to be the smallest value satisfying the following conditions:
	\begin{itemize}
		\item[(iii)] there are no arcs $(v,u)\in E$ with $v\in V_B\cup V_R$, $x(v)\leq t_u$ and $x(u)>t_u$;
		\item[(iv)] there are no arcs $(v,u)\in\EXT_x$ with $v\in V_W$, $x(v)\leq t_u$ and $x(u)>t_u$,
	\end{itemize}
	and let $Y=\{v\in V \mid x(v)\le t_u\}$. Note that both $t_l$ and $t_u$ trivially exist. 
	Note also that the sets $X$ and $Y$ can be computed in $O(|V|\log|V|+|E|)$ time.
	
	\begin{lemma}\label{l2-}
		It holds that $\max\{-t_l,t_u-x^-\}\leq nR_hD^{k}$.
	\end{lemma}
	
	\noindent{To prove Lemma~\ref{l2-}, we need the following lemma.}
	\begin{lemma}\label{l1-}
		Consider any move $(v,u)\in E$ and let $x$ be the current potential. Then
		$$
		x(u)\geq
		\left\{
		\begin{array}{ll}
		x(v)-(m_h^+-r_h^-)&\mbox{ if either $(v\in V_W$  and $(v,u)\in E)$ or $(v\in V_B$ and $(v,u)\in\EXT_x)$}\\
		D\left[x(v)-(m_h^+-r_h^-)\right]&\mbox{ if $v\in V_R$ and $(v,u)\in E$},
		\end{array}
		\right.
		$$
		and
		$$
		x(u)\leq
		\left\{
		\begin{array}{ll}
		x(v)+r_h^+-m_h^-&\mbox{ if either  $(v\in V_B$  and  $(v,u)\in E)$ or $(v\in V_W$ and $(v,u)\in\EXT_x)$}\\
		D\left[x(v)+r_h^+-m_h^--(1-\frac{1}{D})x^-\right]&\mbox{ if $v\in V_R$ and $(v,u)\in E$}.
		\end{array}
		\right.
		$$
	\end{lemma}
	
	\proof
	We only consider the case for $v\in V_R$, as
	the other claims are obvious from the definitions.
	For the first claim, assume that $x(v)\ge x(u)$, since otherwise there is nothing to prove. Then
	from $m_x(v)\leq m_h^+$, it follows that
	\begin{eqnarray*}
		m_h^+-r_h^-&\geq& m_h^+-\sum_{u'}p(v,u')r_h(v,u')\\
		&\geq& p(v,u)(x(v)-x(u))+\sum_{u'\neq u}p(v,u')(x(v)-x(u'))\\
		&\geq& \frac{1}{D}(x(v)-x(u))+x(v)(1-\frac{1}{D}),
	\end{eqnarray*}
	where the last inequality follows from the non-positivity of $x$. This proves the first claim. The other claim can be proved by a similar argument (by replacing $x(\cdot)$ by $x^--x(\cdot)$ and $m^+_h-r_h^-$ by $r_h^+-m_h^-$).
	\qed
	
	\medskip
	
	\noindent{\bf Proof of Lemma \ref{l2-}.}~
	By definition of $X$, for every position $v\in X$ there must exist (not necessarily distinct) positions $v_0,v_1,\ldots, v_{2j}=v\in X$, 
	$j\le |X|$, such that $x(v_0)=0$, and for $i=1,2,\ldots,j$, $x(v_{2i})\ge x(v_{2i-1})$, 
	and either ($(v_{2i-2},v_{2i-1})\in E$ and $v_{2i-2}\in V_W\cup V_R$) or 
	($(v_{2i-2},v_{2i-1})\in\EXT_x$ and $v_{2i-2}\in V_B$). 
	Among the even-numbered positions, let $v_{2i_1-2},\ldots,v_{2i_l-2}$ be the ones belonging to $V_R$, and assume withut loss of generality that $l>0$ and $i_1<i_2<\cdots<i_l$.
	Using Lemma \ref{l1-}, we obtain the following inequality by a telescoping sum:
	\begin{equation}\label{ee1}
	x(v_{2i_{q+1}-2})\geq x(v_{2i_{q}-1})-(i_{q+1}-i_q-1)(m^+_h-r_h^-),\mbox{ for }q=1,\ldots,l-1,
	\end{equation}
	and $x(v_{2i_1-2})\geq -(i_1-1)(m^+_h-r_h^-)$.
	
	Now applying Lemma \ref{l1-} to the pair $v_{2i_q-2}\in V_R$ and $v_{2i_q-1}$, for $q=1,\ldots,l-1$, and using \raf{ee1} we obtain:
	\begin{equation}\label{ee2}
	x_{q+1}\ge D x_q-(D+i_{q+1}-i_q-1)(m^+_h-r_h^-),~~~~x_1\ge-(i_1-1)(m^+_h-r_h^-), 
	\end{equation}
	where we write, for convenience, $x_q=x(v_{2i_q-2})$, for $q=1,\ldots,l$.  Iterating, we obtain:
	$$
	x_l\geq -\left(D^{l-1}(i_1-1)+\sum_{q=2}^{l}D^{l-q}(D+i_q-i_{q-1}-1)\right)(m^+_h-r_h^-).
	$$
	Combining this with the inequality $x(v)\geq D x_l-(D+j-i_l)(m^+_h-r_h^-)$ and using $D>1$, we get
	$$
	x(v)\geq -D^{l} j(m_h^+-r^-_h)\geq -D^{k}|X|(m_h^+-r^-_h).
	$$
	Similarly, one can prove for any $v\in Y$ that $x(v)\leq x^-+D^{k}|Y|(r_h^+-m_h^-)$, and the lemma follows.
	\qed
	
	\medskip
	
	The correctness of the algorithm follows from the following lemma.
	
	\begin{lemma}\label{l-gap}
		Suppose that pumping is performed for $N_h\geq 2nT_h+1$ iterations, where
		$T_h=\frac{4nR_hD^{k}}{M_h}$, and neither the set $V_x[t_0,t_1)$ nor $V_x(t_3,t_4]$ becomes empty. Let $V^-=X$ and $V^+=Y$ be the sets constructed as above, and $V^0=V\setminus (X\cup Y)$. Then $V^+ \cup V^- \cup V^0$ is a contra-ergodic decomposition.
	\end{lemma}
	\proof 
	We pump in each iteration by $\delta\geq\frac{M_h}{4}$.
	Furthermore, our formula for $\delta$ implies that once a vertex enters the region $V_x[t_1,t_3]$, 
	it never leaves this region. 
	In particular, there are vertices $v_0\in X\cap V_x[t_0,t_1)$ and $v_n\in Y\cap V_x(t_3,t_4]$ with $x(v_0)=0$ and $x(v_n)=x^-$.
	
	For a vertex $v\in V$, let $N(v)$ denote the number of times the vertex was pumped. Then $N(v_0)=0$ and $N(v_n)=N_h$.
	
	We claim that $N(v)\leq T_h$ for any $v\in X$, and $N(v)\ge N_h-T_h$ for any $v\in Y$ 
	(i.e., every vertex in $X$ was \emph{not} pumped in \emph{all} steps but at most $T_h$, 
	and every vertex in $Y$ was pumped in \emph{all} steps but at most $T_h$). 
	Indeed, if $v\in X$ (respectively, $v\in Y$) was pumped greater than (respectively, less than) $T_h$ times then
	$x(v)-x(v_0)\le -nR_hD^{k}$ (respectively, $x(v_n)-x(v)\le -nR_hD^{k}$), in contradiction to Lemma \ref{l2-}.
	
	Since $N_h>2T_h$, it follows that $X\cap Y=\emptyset$. Furthermore, among the first $2nT_h+1$ iterations, 
	in at most $nT_h$ iterations some vertex $v\in X$ was pumped, and in at most $nT_h$ iterations some vertex in $Y$ was not pumped. 
	Thus, there must exist an iteration at which every vertex $v\in X$ was not pumped and every vertex $v\in Y$ was pumped. 
	At that particular iteration, we must have $X\subseteq V_x[t_0,t_2)$ and $Y\subseteq V_x[t_2,t_4]$, 
	and hence $m_x(v)< t_2$ for every $v\in X$ and $m_x(v)\ge t_2$ for every $v\in Y$. 
	By the way the sets $X$ and $Y$ were constructed, 
	we can easily see that $X$ and $Y$ will continue to have this property till the end of the $N_h$ iterations, and hence they induce a contra-ergodic partition. The lemma follows.
	\qed
	
	\subsection{Potential reduction: REDUCE-POTENTIALS$(\cG,x)$}
	\label{sec:pot-red}
	One problem that arises during the pumping procedure is that the potentials can increase exponentially in the number of phases, 
	making our bounds on the number of iterations per phase also exponential in $n$. 
	For the BW-case Pisaruk \cite{Pis99} solved this problem by giving a procedure that reduces the range of the potentials after each round,
	while keeping all its desired properties needed for the running time analysis. 
	
	Pisaruk's potential reduction procedure can be thought of as a combinatorial procedure for finding an \emph{extreme point} of a polyhedron, 
	given a point in it.
	Indeed, given a BWR-game and a potential $x$, let us assume without loss of generality, by shifting the potentials if necessary, 
	that $x\geq 0$, and let $E'=\{(v,u)\in E\mid r_x(v,u)\in[m_x^-,m_x^+], ~v\in V_B\cup V_W\}$, 
	where $r$ is the \emph{original} local reward function. 
	Then the following polyhedron
	{\small
		\[
		\Gamma_x=\left\{x'\in \RR^V\left|
		\begin{array}{cl}m_x^- \leq  r(v, u) + x'(v) - x'(u)\leq m_x^+,& \forall (v, u) \in E'\\*[3mm]
		r(v, u) + x'(v) - x'(u)\leq m_x^+,& \forall v\in V_W,~(v, u) \in E\setminus E'\\*[3mm]
		m_x^-\leq r(v, u) + x'(v) - x'(u),& \forall v\in V_B,~(v, u) \in E\setminus E'\\*[3mm]
		m_x^-\leq \sum_{u \in V} p(v, u)(r(v,u)+x'(v)-x'(u))\leq m_x^+,&\forall v \in V_R\\*[3mm]
		x'(v)\ge 0 &\forall
		v\in V\end{array}\right.\right\}.
		\]}
	
	\noindent
	is non-empty, since $x\in\Gamma_x$.
	Moreover, $\Gamma_x$ is pointed, and hence, it must have an extreme point.
	Let us remark that given a feasible point $x$, 
	an extreme point can be computed in $O(n^2|E|)$ time (see, e.g., \cite{Scr03}).
	
	\begin{lemma}\label{pot-red}
		Consider a BWR-game in which all rewards are integral with range $R=r^+-r^-$, 
		and probabilities $p(v,u)$ are rational with common denominator $D$, 
		and let $k=|V_R|$. Then any extreme point $x^*$ of $\Gamma_x$ satisfies $\|x^*\|_{\infty}\leq n Rk(2D)^{k}$.
	\end{lemma}
	\proof
	Consider such an extreme point $x^*$.
	Then $x^*$ is uniquely determined by a system of $n$ linearly independent equations chosen from the given inequalities. 
	Thus there exist subsets $V'\subseteq V$, $V_R'\subseteq V_R$ and $E''\subseteq E$ such that $|V'|+|V_R'|+|E''|=n$, 
	$x^*$ is the unique solution of the subsystem $x'(v)=0$ for all $v\in V'$, 
	$x'(v)-x'(u)=m^*_x-r(v,u)$ for $(v,u)\in E''$, 
	and $x'(v)-\sum_{u \in V} p(v, u)x'(u)=m^*_x-\sum_{u \in V} p(v, u)r(v,u)$ for $v\in V_R'$, 
	where $m^*_x$ stands for either $m^-_x$ or $m^+_x$.
	
	\medskip
	
	Note that all variables $x'(v)$ must appear in this subsystem, 
	and that the underlying undirected graph of the digraph $G'=(V,E'')$ must be a forest
	(otherwise the subsystem does not uniquely fix $x^*$, or it is not linearly independent).
	
	Consider first the case $V_R=\emptyset$. For $i\geq 0$, 
	let $V_i$ be the set of vertices of $V$ at (undirected) distance $i$ from $V'$ (observe that $i$ is finite for every vertex). 
	Then we claim by induction on $i$ that
	$x^*(v)\leq i\gamma$ for all $v\in V_i$, where $\gamma=\max\{m_x^+-r^-,r^+-m_x^-\}$. This is trivially true for $i=0$. 
	So let us assume that it is also true for some $i>0$.
	For any $v\in V_{i+1}$, there must exist either an arc $(v,u)$ or an arc $(u,v)$ where $u\in V_{i-1}$. 
	In the former case, we have $x^*(v)=x^*(u)+m^*_x-r(v,u)\leq i\gamma+m_x^+-r^-\leq (i+1)\gamma$. 
	In the latter case, we have $x^*(v)=x^*(u)-(m^*_x-r(u,v))\leq i\gamma+r^+-m_x^-\leq (i+1)\gamma$.
	
	Now suppose that $|V_R|>0$. For each connected component $D_l$ in the forest $G'$, 
	let us fix a position $u_l$ from $V'$ if  $D_l\cap V'\neq\emptyset$; otherwise, $v_l$ is chosen arbitrarily.
	For every position $v\in D_l$ let $\cP_v$ be a (not necessary directed) path from $v$ to $v_l$. 
	Thus, we can write $x'(v)$ uniquely as
	
	\begin{equation}\label{express}
	x'(v)=x'(v_l)+\ell_{v,1}m_x^++\ell_{v,2}m_x^-+\sum_{(u',u'')\in \cP_v}\ell_{v,u',u''}r(u',u''),
	\end{equation}
	for some $\ell_{v,1},\ell_{v,2}\in\ZZ$, and $\ell_{v,u',u''}\in\{-1,1\}$.
	Thus if $x^*(v_l)=0$ for some component $D_l$, 
	then by a similar argument as above, $x^*(v) \leq \gamma|D_l|$ for every $v\in D_l$.
	
	Note that, up to this point, we have used all equations corresponding to arcs in $G'$ and to vertices in $V'$. 
	The remaining set of $|V_R'|$ equations should uniquely determine the values of the variables in any component which has no position in $V'$.
	Substituting the values of $x'(v)$ from \raf{express}, for the positions in any such component, 
	we end-up with a linearly independent system on $k'=|V_R'|$ variables $Ax=b$, where
	$A$ is a $k'\times k'$ matrix in which eatch entry is at most 1 in absolute value 
	and the sum of each row is  at most 2 in absolute value, and $\|b\|_{\infty}\leq n(R+M_x)\leq 2nR$.
	
	The rest of the argument follows (in a standard way) by Cramer's rule. 
	Indeed, the value of each component in the solution is given by $\Delta'/\Delta$, 
	where $\Delta$ is the determinant of $A$ and $\Delta'$ is the determinant of a matrix  
	obtained by replacing one column of $A$ by $b$. 
	We upper bound $\Delta'$ by $k'\|b\|_{\infty}\Delta_{max}$, 
	where $\Delta_{max}$ is the maximum absolute value of a subdeterminant of $A$ of size $k'-1$. 
	To bound $\Delta_{max}$, let us consider such a subdeterminant with rows $a_1,\ldots, a_{k'-1}$, 
	and use Hadamard's inequality:
	\begin{equation*}\label{Num}
	\Delta'\leq \prod_{i=1}^{k'-1} \|a_i\|\leq 2^{k'-1},
	\end{equation*}
	since $\|a_i\|_1\le 2$, for all $i$. To lower bound $\Delta$, we note that $D^{k'}\Delta$ is a non-zero integer, and hence has absolute value at least 1. Combining the above inequalities, the lemma follows.
	\qed

	Note that any point $x'\in \Gamma_x$ satisfies $M_{x'} \subseteq M_{x}$, 
	and hence replacing $x$ by $x^*$ does not increase the range of $m_x$.

	\subsection{Running time analysis}\label{ss21}
	From Lemmas \ref{l-gap} and \ref{pot-red}, we can conclude the following bound.
	\begin{lemma}\label{l-run}
		Procedure PUMP$(\cG,\eps)$ terminates in 
		$O(n^2|E|R( n k(2D^2)^k\frac{1}{\eps}+ \log R)$ time.
	\end{lemma}
	
	\proof
	We note the following:
	\begin{enumerate}
		\item By \raf{N}, the number of iterations per phase $h$ is at most $N_h=\left\lfloor\frac{8n^2R_hD^k}{M_{h}}\right\rfloor+1$.
		\item Each iteration requires $O(|E|)$ time, 
		and the end of a phase we need  additional $O(n^2|E|)$ time  
		(which is required for REDUCE-POTENTIALS).
		\item By Lemma \ref{pot-red}, for any $(v,u)\in E$, 
		we have $r_x(v,u)=r(v,u)+x(v)-x(u)\leq  r(v,u)
		+2nk(2D)^kR$, 
		and similarly, 
		$r_x(v,u)\geq r(v,u) -2nk(2D)^kR.$ 
		In particular, $R_h\leq (1+4nk(2D)^k)R$ at the beginning of each phase $h$ in the procedure.
	\end{enumerate}
	
	Since $M_h \leq \frac{3}{4}M_{h-1}$ for $h=1,2,\ldots$, 
	the maximum number of such phases until 
	we reach the required accuracy is at most $H=\log_{4/3}\left(\frac{M_0}{\eps}\right)$.
	Putting all the above together, we get that the total running time is at most
	$$
	\sum_{h=0}^H\frac{8n^2D^k|E|(1+4nk(2D)^k)R}{M_h}+O(n^2|E|)H.
	$$
	Noting that $M_0 \leq R$ and $M_H \geq \eps$, the lemma follows.
	\qed
	
	We now give an upper bound on the required accuracy $\eps$.
	%
	
	\begin{lemma}\label{l1-prob}
		Consider a Markov chain $\cM=(G=(V,E),P)$ with $n$ positions among which $k$ are random and in which all the entries of the transition matrix $P$ are rational numbers with common denominator $D$. Assume that $\cM$ has only one absorbing class. Let $p^*$ be the limiting distribution starting form any position. 
		Then $p^*(v)$, for all positions $v$, are rational numbers with common denominator at most $(k+1)n(2D)^{k+3}$. 
	\end{lemma}
	\proof
	For $v\in V$, let $\cR(v)$ be the set of positions that can reach $v$ by a directed path in $G$, all whose internal positions (if any) are in $V_B\cup V_W$.  
	Note that $\cR(v)\cap V_R\neq\emptyset$ for each $v\in V_B\cup V_W$.
	Otherwise, there is a position $v\in V_B\cup V_W$ such that no random position in $V$ can reach $v$, implying by the strong connectivity of $G$ that $V_R=\emptyset$, and hence $\cM$ must be a Hamiltonian cycle on deterministic positions with $p^*(v)=\frac{1}{n}$ for all $v\in V$.
	
	Consider the system of equations in (L5) defining $p^*(v)$: 
	$$p^*(v)=\sum_{u\in V_B\cup V_W: (u,v)\in E}p^*(u)+\sum_{u\in V_R}p(u,v)p^*(u),$$ for $v\in V$,
	and $\sum_{v\in V}p^*(v)=1$. 
	
	Eliminating the variables $p^*(v)$ for $v\in V_B\cup V_W$:
	\begin{equation}
	\label{e1.1.1}
	p^*(v)=\sum_{u\in \cR(v)\cap V_R}p'(u,v)p^*(u),
	\end{equation}
	where $p'(u,v)=p(u,v)+\sum_{u'\in\cR(v)\cap (V_B\cup V_W)}p(u,u')$,
	we end-up with a system on only random positions $v\in V_R$: $p^*(v)=\sum_{u\in V_R}p'(u,v)p^*(u)$. (Note that $\sum_{u\in V_R}p'(v,u)=1$ for all $v\in V_R$.) 
	Similarly, we can reduce the normalization equation to $\sum_{v\in V_R}(1+\sum_{u\in V_B\cup V_W:v\in\cR(u)}p'(v,u))p^*(v)=1$.
	This gives a system on $k$ variables of the form $(p^*)^T(I-P')=0,~(p^*)^Tb=1$, 
	where the matrix $P'$ and the vector $b$ have rational entries with common denominator $D$, each row of $P'$ sums up to $1$, and $b_i$ is rational number $\in[1,n]$ with denominator $D$.
	
	Let us multiply by $D$ all the equations of this system and note that all coefficients of the resulting system of equations are integers in $[-D,D]$ for the first $k-1$ equations, and in $[D,nD]$ for the normalization equation.
	Any non-zero component $p^*(v)$ in the solution of this system takes the form $\frac{D\Delta_0}{\sum_{i=1}^{k}Db_i\Delta_i}$, 
	where $\Delta_0,\Delta_1,\ldots,\Delta_{k}$ are subdeterminants of $D(I-P')$ of rank $k-1$. 
	It follows by Hadamard's inequality that $\Delta_i\le (2D)^{k}$, and hence $\sum_{i=1}^{k}Db_i\Delta_i$ is an integer of value at most $kn(2D)^{k+2}$.
	
	After solving this system, we can get the value of $p^*(v)$, for $v\in V_W\cap V_B$, from \raf{e1.1.1} as rational numbers of common denominator at most $kn(2D)^{k+3}$. 
	\qed
	
	\begin{theorem}\label{cor1}
		When procedure PUMP$(\cG,\eps)$ is run with $\eps$ as in \raf{accur}, 
		it either outputs a potential vector $x$ such $m_x(v)$ is constant for all $v\in V$, 
		or finds a contra-ergodic partition. 
		The total running time is $\poly(n)(2D)^{O(k)}R\log R$.
	\end{theorem}
	\proof
	Suppose that the game is not ergodic and let us fix an optimal situation (pair of optimal strategies) $s$. Then there must exist at least two absorbing classes in the obtained weighted Markov chain which have different values. Consider such an absorbing class and contract all arcs $(v,u)$ where $v\in V_W\cup V_B$. We obtain an absorbing Markov chain on a subset of $V_R$. We now can apply Lemma \ref{l1-prob} to conclude that the value of any position in this class is a rational number with  denominator at most $(k+1)n(2D)^{k+3}$. Consequently, the difference between any two different values of absorbing classes of $\cG_s$ is at least $((k+1)^2n^2(2D)^{2k+6})^{-1}$. If PUMP$(\cG,\eps)$ terminates with $M_x<\eps$ then we conclude by Lemma \ref{l1-prob}, analogously to the previous arguments, that all $m_x(v)$ values are the same. This implies by Proposition~\ref{p1-} that this is the optimal value and the locally optimal moves are globally optimal. The running time bound follows now from Lemma \ref{l-run}.
	\qed
	
	\subsection{Lower bound for ergodic games}
	\label{Exp-ex}
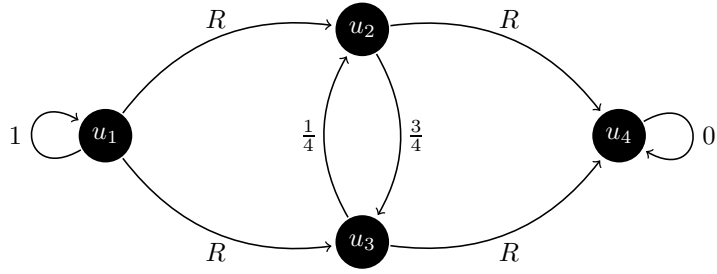
\begin{figure}[htb!]
		
		\begin{tikzpicture}[->,shorten >=1pt,auto,node distance=2cm,semithick]
		\tikzstyle{every node}=[draw,shape=circle]
		
		\node[white,fill=black] (u1) {$u_1$};
		\node[draw=none] (v1) [right of=u1] {};
		\node[white,fill=black] (u2) [above right of =v1] {$u_2$};
		\node[white,fill=black] (u3) [below right of =v1] {$u_3$};
		\node[draw=none] (v2) [above right of =u3] {};
		\node[white,fill=black] (u4) [right of =v2] {$u_4$};
		
		\path[->,min distance=1cm]
		(u4) edge [right,in=-30,out=30] node[draw=none,right=-1mm] {$0$} (u4)
		(u1) edge [left,in=150,out=210] node[draw=none,left=-1mm] {$1$} (u1);
		
		\path[->]
		(u1) edge [bend left] node[draw=none,above=-1mm] {$R$} (u2)
		(u1) edge [bend right] node[draw=none,below=-1mm] {$R$} (u3)
		(u2) edge [bend left] node[draw=none,above=-1mm] {$R$} (u4)
		(u3) edge [bend right] node[draw=none,below=-1mm] {$R$} (u4)
		(u2) edge [bend left] node[draw=none,right=-2mm] {$\frac{3}{4}$} (u3)
		(u3) edge [bend left] node[draw=none,left=-2mm] {$\frac{1}{4}$} (u2);

		\end{tikzpicture}
		\centering
		\caption{A lower bound example for pumping algorithm for B-games. In this example positions $u_2$ and $u_3$ will oscillate $2R$ times before converging to value $0$. Labels attached to the arcs represent rewards.}
		\label{f2}
	\end{figure}

	Note that if $|V_R|=0$, there is a simple example with only one player showing that the running time of the pumping algorithm can be linear in $R$; see Figure~\ref{f2}.
	
\begin{figure}[htb!]
		\begin{tikzpicture}[->,shorten >=1pt,auto,node distance=2cm,semithick]
		\tikzstyle{every node}=[draw,shape=circle]
		
		\node[fill=red!20] (u0)  {$u_0$};
		\node (u1) [left of =u0] {$u_1$};
		\node (u2) [left of =u1] {$u_2$};
		\node (u3) [left of =u2] {$u_3$};
		\node (v1) [right of =u0] {$v_1$};
		\node (v2) [right of =v1] {$v_2$};
		\node (v3) [right of =v2] {$v_3$};  
		
		\path[->,min distance=1cm]
		(v3) edge [right,in=-30,out=30] node[draw=none,right=-1mm] {$\frac{D-1}{D}$} (v3)
		(u3) edge [left,in=150,out=210] node[draw=none,left=-1mm] {$\frac{D-1}{D}$} (u3);
		
		\path
		(u3) edge[bend left] node[draw=none,above=-1mm] {$\frac{1}{D}$} (u2)
		(u2) edge[bend left] node[draw=none,above=-1mm] {$\frac{1}{D}$} (u1)
		(u1) edge[bend left] node[draw=none,above=-1mm] {$\frac{1}{D}$} (u0)
		(u0) edge[bend left] node[draw=none,above=-1mm] {$\frac{1}{2}$} (v1)
		(v1) edge[bend left] node[draw=none,above=-3mm] {$\frac{D-1}{D}$} (v2)
		(v2) edge[bend left] node[draw=none,above=-3mm] {$\frac{D-1}{D}$} (v3);
		
		\path
		(v3) edge[bend left] node[draw=none,below=-1mm] {$\frac{1}{D}$} (v2)
		(v2) edge[bend left] node[draw=none,below=-1mm] {$\frac{1}{D}$} (v1)
		(v1) edge[bend left] node[draw=none,below=-1mm] {$\frac{1}{D}$} (u0)
		(u0) edge[bend left] node[draw=none,below=-1mm] {$\frac{1}{2}$} (u1)
		(u1) edge[bend left] node[draw=none,below=-3mm] {$\frac{D-1}{D}$} (u2)
		(u2) edge[bend left] node[draw=none,below=-3mm] {$\frac{D-1}{D}$} (u3);
		
		\end{tikzpicture}
		\caption{An exponential time example. Labels attached to the arcs represent transition probabilities. Local rewards on all arcs are $0$ except on the two loops, where $r(u_3,u_3)=1$ and $r(v_3,v_3)=-1$.}
		\label{f1}
	\end{figure}
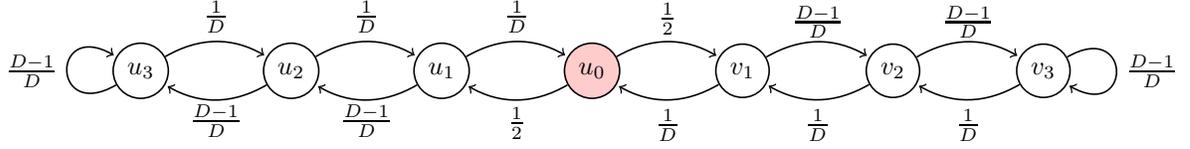

	We show now that the execution time of the algorithm, in the worst case,
	can be exponential in the number of random positions $k$, already for weighted
	Markov chains, that is, for R-games.  Consider the following example; see Figure~\ref{f1}.
	Let $G=(V,E)$ be a digraph on $k=2l+1$ vertices $u_l,\ldots, u_1,u_0=v_0,v_1,\ldots,v_l$, and with the
	following set of arcs:
	$$
	E=\{(u_l,u_l),(v_l,v_l)\}\cup\{(u_{i-1},u_i),(u_i,u_{i-1}),(v_{i-1},v_i),(v_i,v_{i-1})\mid i=1,\ldots, l\}.
	$$
	Let $D\geq 2$ be an integer. All positions are random with the following transition probabilities:
	$p(u_l,u_l)=p(v_l,v_l)=1-\frac{1}{D}$,
	$p(u_0,u_1)=p(u_0,v_1)=\frac{1}{2}$,
	$p(u_{i-1},u_i)=p(v_{i-1},v_i)=1-\frac{1}{D}$, for $i=2,\ldots,l$, and
	$p(u_i,u_{i-1})=p(v_i,v_{i-1})=\frac{1}{D}$, for $i=1,\ldots,l$.
	The local rewards are zero on every arc, except for $r(u_l,u_l)=-r(v_l,v_l)=1$ 
	(See Figure \ref{f1} for $l=3$).
	Clearly this Markov chain consists of a single recurrent class, and it is easy to verify that
	the limiting distribution $p^*$ is as follows:
	\begin{eqnarray*}
		p^*(u_0)&=&\frac{D-2}{D(D-1)^l-2},~
		p^*(u_i)=p^*(v_i)=\frac{(D-1)^{i-1}((D-1)^2-1)}{2D(D-1)^l-2}\mbox{ for $i=1,\ldots,l$}.
	\end{eqnarray*}
	The optimal expected reward at each vertex is
	$$
	\mu(u_i)=\mu(v_i)=-1\cdot(1-\frac{1}{D})p^*(u_l)+1\cdot(1-\frac{1}{D})p^*(u_l)=0,
	$$
	for $i=0,\ldots,l$. Up to a shift, there is a unique set of potentials $x$ that
	transform the Markov chain into the canonical form, 
	and they satisfy the following system of equations:
	\begin{eqnarray*}
		0&=&-\frac{1}{2}\Delta_1-\frac{1}{2}\Delta_1',\\
		0&=&-(1-\frac{1}{D})\Delta_{i+1}+\frac{1}{D}\Delta_i,\mbox{ for }i=1,\ldots, k-1\\
		0&=&-(1-\frac{1}{D})\Delta_{i+1}'+\frac{1}{D}\Delta_i',\mbox{ for }i=1,\ldots, k-1\\
		0&=&-(1-\frac{1}{D})+\frac{1}{D}\Delta_l,\\
		0&=&1-\frac{1}{D}+\frac{1}{D}\Delta_l',\\
	\end{eqnarray*}
	where $\Delta_i=x(u_i)-x(u_{i-1})$ and $\Delta_i'=x(v_i)-x(v_{i-1})$;
	by solving the system, 
	we get $\Delta_i=-\Delta_i'=(D-1)^{k-i+1},$ for $i=1,\ldots,l$.
	
	\smallskip
	
	\noindent{\bf Lower bound on pumping algorithms.}~ Any pumping algorithm that starts with  $0$ potentials and modifies the potentials in each iteration
	by at most $\gamma$ will not have a number of iterations less than $\frac{(D-1)^{l-1}}{2\gamma}$ on the above example. 
	In particular, the algorithm in Section \ref{ErgodPump}
	has $\gamma\le 1/\min\{p(v,u)\mid (v,u)\in E,~p(v,u)\neq 0\}$, which is $\Omega(D)$ in our example. 
	We conclude that the running time of the algorithm is $\Omega(D^{l-2})=D^{\Omega(k)}$ on this example.
	
	\section{Non-ergodic BWR-games}\label{sec:non-ergodic}
	The main difficulty in solving non-ergodic BWR-games is the fact that random positions may introduce transient classes. In fact, our algorithm developed for the ergodic case can be used to find the top/bottom classes, and if no random node enters these classes from outside, then we could call the same procedure (called Find-Top$(\cdot)$/Find-Bottom$(\cdot)$) recursively for the rest of the positions and find an optimum solution. However, this is not always the case. To handle this we need to introduce parametrized games.
	
	\subsection{Preliminary results and basic lemmas}\label{sec:basic}
	Let $\cG=(G=(V=V_B\cup V_W\cup V_R,E),p,r)$ be a BWR-game. In what follows we will use the following notation.
	For a $\theta\in\RR$, let $S(\theta):=\{v\in V~|~\mu_{\cG}(v)=\theta\}$. If $S(\theta)\neq\emptyset$, we refer to it as an {\it ergodic class}. If $S(\theta)=V$ the game $\cG$ is said to be ergodic. Every BWR-game $\cG$ has a unique sequence of numbers $\theta_1<\theta_2<\cdots<\theta_\ell$ such that $S(\theta_i)\neq\emptyset$ for all $i\in[\ell]$ and $\bigcup_{i=1}^\ell S(\theta_i)=V$. 
	
	
	\begin{proposition}\label{p1}
		Ergodic classes necessarily satisfy the following properties.
		
		\begin{itemize}
			\item[(i)] There exists no arc $(v,u)\in E$ such that $v\in V_W\cap S(\theta_i)$, $u\in S(\theta_j)$, and $j>i$;
			\item[(ii)] there exists no arc $(v,u)\in E$ such that $v\in V_B\cap S(\theta_i)$, $u\in S(\theta_j)$, and $j<i$;
			\item[(iii)] for every $v\in V_W\cap S(\theta_i)$, there exists an arc $(v,u)\in E$ such that $u\in S(\theta_i)$;
			\item[(iv)] for every $v\in V_B\cap S(\theta_i)$, there exists an arc $(v,u)\in E$ such that $u\in S(\theta_i)$;
			\item[(v)] there exists no arc $(v,u)\in E$ such that $v\in V_R\cap S(\theta_1)$, and $u\not\in S(\theta_1)$;
			\item[(vi)] there exists no arc $(v,u)\in E$ such that $v\in V_R\cap S(\theta_\ell)$, and $u\not\in S(\theta_\ell)$.
		\end{itemize}
	\end{proposition}
	\proof
	All claims follow from the existence of canonical form for $\cG$ by Theorem \ref{t2}, since the existence of arcs forbidden by (i), (ii), (v), or (vi), or the non-existence of arcs required by (iii) and (iv), would violate the value equations (C1).
	\qed
	
	For a set of positions $S\subseteq V$, we define the {\it black closure} $\Cl_B(S)$ (respectively, the {\it black semi-closure} $\Cl_B'(S)$) of $S$  to be the set of positions which is recursively obtained from $S$ by adding
	\begin{itemize}
		\item [(1)] a position $v \in  V_B$ (respectively, $v \in  V_B \cup V_R$), if some arc $(v,u) \in E$ satisfies $u \in S$,
		or
		\item[(2)] a position $v \in V_W \cup V_R$ (respectively, $v \in  V_W$), if all arcs  $(v,u) \in E$ satisfy
		$u \in S$.
	\end{itemize}
	In words, $\Cl_B(S)$ (respectively, $\Cl_B'(S)$) is the set of positions to which \Min\ can force a move with probability 1 (respectively, with some positive probability).
	
	The white closure and semi-closure of $S$, $\Cl_W(S)$ and $\Cl_W'(S)$, are defined analogously. Associated with a black closure (respectively white closure $\Cl_W(S)$) is a (partial) strategy $s_B(\Cl_B(S))$ (respectively, $s_W(\Cl_W(S))$) which guarantees \Min\ (respectively, \Max) a move into $S$. Similar strategies are defined with respect to semi-closures.
	
	\begin{observation}\label{Ob1}
		If $\mu_\cG(v)<z$ for all $v\in S$ then the same holds for all $v\in\Cl_B(S)$. Furthermore, if $z=\theta_\ell$ then the same holds for all $v\in\Cl'_B(S)\supseteq\Cl_B(S)$.  
	\end{observation}
	
	We now introduce parametrized games.
	
	\medskip
	
	To a BWR-game $\cG$ and a set $Y\subseteq V$ such that every $v\in V_W\cup V_B$ has some arc $(v,u)$ with $u\in Y$, we associate the restriction $\cG[Y]$ obtained by deleting all positions in $V\setminus Y$ and all arcs $(v,u)$ where $v\not\in Y$ or $u\not\in Y$. Note that $\cG[Y]$ is not a BWR-game in general.  Given such a restriction and $x\in\RR$, we further define a BWR-game $\cG[Y](x)$ and a BW-game $\widehat{\cG[Y]}(x)$ as follows:
	\begin{itemize}
		\item [{$\cG[Y](x)$}:] we add a new deterministic position $w=w(Y)$ with a self-loop $(w,w)$ with local reward $x$, and for every $v\in V_R\cap Y$ add an arc from $v$ to $w(Y)$ with local reward $0$ and transition probability $p(v,w(Y)):=1-\sum_{u\in Y}p(v,u)$ if $p(v,w(Y))>0$; 
		
		\item[{$\widehat{\cG[Y]}(x)$}:] we remove all arcs leaving $v\in V_R\cap Y$ and contract the set $v\in V_R\cap Y$ into a deterministic position $w(Y)$ (black or white, arbitrarily) with a self-loop $(w(Y),w(Y))$ having local reward value $x$. 
	\end{itemize}
	We call such games {\it parametrized} BWR-/BW-games (with parameter $x$).
	
	For a BW-game $\cG$, let $\nu(\cG)$ be the number of distinct optimal values achieved by positions of $\cG'$. For a BWR-game $\cG$, we extend this by defining $\nu(\cG):=\max_{Y\subseteq V,~x\in\RR}\nu(\widehat{\cG[Y]}(x))$. 
	
	For a situation $s=(s_W,s_B)$ such that $s(v)\in Y$ for all $v\in Y$, we denote by $s[Y]:=(s_W[Y],s_B[Y])$ the restriction of $s$ on $Y$.
	For brevity in what follows, we will call $(v,u)$ a black, white, or random arc, depending on whether $v\in V_B$, $v\in V_W$, or $v\in V_R$, respectively.
	
	\begin{lemma}\label{l1}
		\begin{itemize}
			\item[(i)] Given a BWR-game $\cG=(G=(V=V_B\cup V_W\cup V_R,E),P,r)$, let $\widehat{\cG}$ be the BW-game obtained from $\cG$ by replacing
			each random position $v\in V_R$ with a terminal deterministic position (black or white, arbitrarily) with a local reward $\mu_{\cG}(v)$ on the
			self-loop $(v,v)$. Then $\mu_{\widehat{\cG}}(v)=\mu_{\cG}(v)$ for all $v\in V$.
			\item[(ii)] Let $\widehat{\cG}$ be as above and $U\subseteq V$ be such that $\mu_{\cG}(v)\ne\mu_{\cG}(u)$ for all $v\in U $ and $u\in V\setminus U$. Then $\mu_{\widehat{\cG}[U]}(v)=\mu_{\cG}(v)$ for all $v\in U$.
		\end{itemize}
	\end{lemma}
	
	\proof
	(i) By Theorem \ref{t2}, there is a potential $x:V\to\RR$ transforming $\cG$ to canonical form and certifying for all positions $v\in V$ that the value of $v$ in $\cG$ is $\mu_{\cG}(v)$. It is obvious that the same potential gives a canonical form for $\widehat\cG$ (namely, given by the canonical form equations for $\cG$, with the equations for the random positions dropped), and hence, by Theorem \ref{t2}  certifies that the value of $v$ in $\widehat\cG$ is also $\mu_{\cG}(v)$. The proof of (ii) follows also form this argument, since $U$ contains some complete ergodic classes of $\cG$.
	\qed

	\medskip
	
	We will write $V(\theta_i):=S(\theta_i)\cup\{w(S(\theta_i))\}$ and denote by  $\bone$ the vector of all ones with appropriate dimension.

	For $i\in[\ell]$, define $\cG[\theta_i]$ to be the game $\cG[S(\theta_i)](\theta_i)$. Proposition \ref{p1} guarantees that the game $\cG[\theta_i]$ is well-defined, that is, for every $v\in S(\theta_i)$ there is at least one arc going out of $v$ in $\cG[\theta_i]$.
	
	The following two lemmas state that if we identify ergodic classes together with their values, then we can find an optimal strategy in the whole game by solving each ergodic class independently. 
	\begin{lemma}\label{l2}
		For all $i\in[\ell]$  and $v\in V(\theta_i)$, it holds that $\mu_{\cG[\theta_i]}(v)=\theta_i$.
	\end{lemma}
	\proof
	Consider a potential transformation $x:V\to\RR$ bringing $\cG$ to canonical form. Let $x':V(\theta_i)\to\RR$ be the vector of potentials defined as follows: $x'(v):=x(v)$ if $v\in S(\theta_i)$, and
	\begin{equation}\label{e1}
	x'(w(S(\theta_i))):=x(v)+\frac{1}{p(v,w(S(\theta_i)))}\left(\sum_{u\in S(\theta_i)}p(v,u)(r(v,u)+x(v)-x(u))-\theta_i\right),
	\end{equation}
	for $v\in V_R\cap S(\theta_i)$.
	Then it is immediate that the pair $(x',\theta_i\bone)$ satisfies the canonical form conditions (C1) and (C2) at any deterministic position $v\in  V(\theta_i)$. Furthermore for $v\in V_R\cap S(\theta_i)$, $(x',\theta_i\bone)$ also satisfies (C1) since, trivially, $\theta_i=\sum_{u\in S(\theta_i)}p(v,u)\theta_i+p(v,w(S(\theta_i)))\theta_i$, and moreover, $$\theta_i=\sum_{u\in S(\theta_i)}p(v,u)(r(v,u)+x(v)-x(u))+p(v,w(S(\theta_i)))(0+x(v)-x'(w(S(\theta_i))))$$ holds by \raf{e1}.
	\qed
	
	\begin{lemma}\label{l3}
		For $i\in[\ell]$, let $s^i:=(s_W^i,s_B^i)$ be a pair of optimal strategies in $\cG[\theta_i]$. Then the situation $s^*=(s_W^*,s_B^*)$ obtained by concatenating all these strategies together (that is, $s^*_W(v):=s^i_W(v)$ for $v\in V_W\cap S(\theta_i)$ and $s^*_B(v):=s^i_B(v)$ for $v\in V_B\cap S(\theta_i)$) is optimal in $\cG$.
	\end{lemma}
	\proof
	For a strategy $s$ denote by $c_s(v)$ the effective payoff \raf{value} starting from position $v$ in the Markov chain $\cG_s$ obtained from $\cG$ by fixing the arcs determined by $s$. The lemma follows from the following claims.
	\begin{Claim}\label{cl1}
		Let $s=(s_W^*,s_B)$ or  $s=(s_W,s_B^*)$, where $s_B\in S_B$ and $s_W\in S_W$ are arbitrary strategies of \Min\ and \Max, respectively.  Then any absorbing class $U$ in the Markov chain $\cG_s$ satisfies: $U\subseteq S(\theta_i)$ for some $i\in[\ell]$.
	\end{Claim}
	\proof
	Without loss of generality, consider $s=(s_W^*,s_B)$ for some $s_B\in S_B$. Suppose that there an absorbing class $U$  in $\cG_s$ such that $U\not\subseteq S(\theta_i)$ for all $i\in[\ell]$. Let $i$ be the largest index such that $U\cap S(\theta_i)\neq\emptyset$. By the strong connectivity of the subgraph induced by $U$ in $\cG_s$, there must exist an arc $(v,u)$ from some $v\in U\cap S(\theta_i)$ to some $u\in U\cap S(\theta_j)\neq\emptyset$, for some $j<i$. By Proposition \ref{p1}-(ii), $v\not\in V_B$, and by the choice of $s$, $v\not\in V_W$. Then $v\in V_R,$ and since $U$ is absorbing, there are no arcs in $G$ from $v$ to some $u\not \in U$, and in particular to no $u\in S(\theta_j)$ with $j>i$. We get the following contradiction from (C1):
	$$
	\theta_i=\sum_{u\in S(\theta_i)}p(v,u)\theta_i+\sum_{u\not\in S(\theta_i)}p(v,u)\mu_{\cG}(u)<\theta_i,
	$$
	since $\mu_{\cG}(u)<\theta_i$ for all $u\not\in S(\theta_i)$.
	\qed
	\begin{Claim}\label{cl2}
		$\mu_{s^*}(v)=\mu_\cG(v)$ for all $v\in V$.
	\end{Claim}
	\proof
	For non-transient positions, the claim follows from Claim \ref{cl1}, which implies that any absorbing class of $\cG_{s^*}$, included in $S(\theta_i)$, is also an absorbing class in the Markov chain $(\cG[\theta_i])_{s^i}$. In particular, the limiting distribution and hence the value of any position $v$ in such an absorbing class are identical to the corresponding ones in $(\cG[\theta_i])_{s^i}$, that is, $\mu_{s^*}(v)=\mu_{\cG[\theta_i]}(v)$. By Lemma \ref{l2}, we get $\mu_{s^*}(v)=\theta_i=\mu_\cG(v)$.
	
	Consider now the set of transient positions $T$. Using the notation in Section~\ref{s-limit}, let $C_1,\ldots,C_h$ be the set of absorbing classes.
	Then it follows from (L3) and (L4) in Section~\ref{s-limit} that $\mu_{s^*}[T]:=(\mu_{s^*}(v):v\in T)$ is the unique solution of the equation
	$
	A\bx=\alpha,
	$
	where $A:=I-P_{s^*}[T;T]$, $\alpha:=\sum_{i=1}^h\mu_iP_{s^*}[T;C_i]\bone$,  and $\mu_i$ is the value $\mu_{\cG}(v)$ for $v\in C_i$. Note that this equation is the value equation given in condition (C1) of the canonical form, where the value $\mu(v)$ is set to $\mu_{\cG}(v)$ for all positions $v$ in the absorbing classes. Since the vector $\bx:=(\mu_{\cG}(v):v\in T)$ satisfies this equation, it is the unique solution, implying that $\mu_{s^*}(v)=\mu_{\cG}(v)$ for all $v\in T$.
	\qed
	\begin{Claim}\label{cl3}
		Let $s'=(s_W^*,s_B)$ and  $s''=(s_W,s_B^*)$, where $s_B\in S_B$ and $s_W\in S_W$ are arbitrary strategies of \Min\ and \Max, respectively.
		Then $\mu_{\cG_{s'}}(v)\ge\mu_{\cG}(v)$ and $\mu_{\cG_{s''}}(v)\le\mu_{\cG}(v)$.
	\end{Claim}
	\proof
	Without loss of generality we only prove the claim for $s'$.
	%
	Let $S\subseteq V$ be the set of absorbing positions in $\cG_{s'}$, and $\xi\in\RR^S$ be the vector of corresponding values. Then the vector of values for the set of transient positions $T:=V\setminus S$ in $\cG_{s'}$ is given by the unique solution of the following equation in $\by$
	\begin{equation}\label{e2'}
	\by=\bar B\left(\begin{array}{c}\by \\ \xi\end{array}\right)=B\by+D\xi,
	\end{equation}
	where $\bar B=\left[\begin{array}{c | c}B&D\end{array}\right]$ is a stochastic matrix with $B:=P_{s'}[T;T]$ and $D:=P_{s'}[T;S]$. Similarly, the value equation for the set of positions $T$ in $\cG_{s^*}$ is given by
	\begin{equation}\label{e2}
	\bx=\bar A\left(\begin{array}{c}\bx \\ \eta\end{array}\right)=A\bx+C\eta,
	\end{equation}
	where $\bar A=\left[\begin{array}{c|c}A&C\end{array}\right]$, $A:=P_{s^*}[T;T]$, $D:=P_{s^*}[T;S]$,
	$\bx\in\RR^T$ and $\eta\in\RR^{S}$ are the vector of values of the positions in $T$ and $S$, respectively.
	
	By Claim \ref{cl2}, $\left[\begin{array}{c|c}\bx&\eta\end{array}\right]=\mu_{\cG}$ satisfies \raf{e2}, and by Claim \ref{cl1} we have $\xi\ge \eta$, since $s_B^i$ is an optimal \Min\ strategy in $\cG[\theta_i]$ for all $i$.
	Note that $\bar A$ an $\bar B$ are identical on any row that corresponds to a position $v\in V_W\cup V_R$. Furthermore, Proposition \ref{p1}-(ii) implies the following {\it shifting property}, for any $v\in V_B$
	\begin{equation*}
	\bar A(v,u)=1 \Longrightarrow \bar B(v,u')=1~~\text{ for some $u'$ such that } \mu_{\cG}(u')\ge  \mu_{\cG}(u),
	\end{equation*}
	which in turn implies that $\bar A\left(\begin{array}{c}\bx \\ \eta\end{array}\right)\le \bar B\left(\begin{array}{c}\bx \\ \eta\end{array}\right) \le \bar B\left(\begin{array}{c}\bx \\ \xi\end{array}\right)$, or
	\begin{equation}\label{e3}
	A\bx+C\eta \le B \bx+D\xi.
	\end{equation}
	By (L3) in Section~\ref{s-limit}, $(I-B)^{-1}$ exists and is non-negative. Combining \raf{e2'}, \raf{e2} and \raf{e3}, we get
	\begin{eqnarray*}
		\bx&\le&(I-B)^{-1}D\xi= \by.
	\end{eqnarray*}
	The claim follows.
	\qed
	
	This completes the proof of Lemma \ref{l3}.
	\qed

	\begin{remark}\label{r1}
		Lemma \ref{l3} states that, if we the know values of all the positions, then we can get uniformly optimal strategies by solving $\ell$ ergodic different games. It should be noted however that, even if we know those values and the corresponding ergodic classes, a pseudo-polynomial algorithm for the ergodic case, as the one described in the previous section, does not yield in general a pseudo-polynomial algorithm for solving the game. The reason is that in our reduction in the proof of Lemma \ref{l3}, we introduce local rewards on self-loops $(v,v)$ of value $\mu_{\cG}(v)$, which might be exponentially small in $n$, even for games with a single random position; see, e.g., \cite{BEGM-ORL}. This is due to the fact that some random positions are transitional, and hence the precision estimate in Lemma \ref{l1-prob} does not apply.
	\end{remark}
	
	In view of the above remark, we need to analyze the structural dependence of the ergodic classes on the guessed values of the random positions. We achieve this by considering a parametrized BW-game as described in the next lemma.
	
	\begin{lemma}\label{AB}
		Let $v$ be a position in a parametrized BW-game $\widehat\cG(x)$, and $x,y\in\RR$.
		\begin{itemize}
			\item[(i)] If $\mu_{\widehat\cG(x)}(v)<x$, then for any $y\ge\mu_{\widehat\cG(x)}(v)$, $\mu_{\widehat\cG(y)}(v)=\mu_{\widehat\cG(x)}(v)$;
			\item[(ii)] if $\mu_{\widehat\cG(x)}(v)>x$, then for any $y\le\mu_{\widehat\cG(x)}(v)$, $\mu_{\widehat\cG(y)}(v)=\mu_{\widehat\cG(x)}(v)$;
			\item[(iii)] if $\mu_{\widehat\cG(x)}(v)=x$, then for any $y>x$, $y\ge\mu_{\widehat\cG(y)}(v)\ge x$;
			\item[(iv)] if $\mu_{\widehat\cG(x)}(v)=x$, then for any $y< x$, $y\le\mu_{\widehat\cG(y)}(v)\le x$.
		\end{itemize}
	\end{lemma}
	\proof
	We prove (i) and (iii); (ii) and (iv) are analogous.
	
	(i) Suppose that $\mu_{\widehat\cG(x)}(v)<x$ and $y\ge\mu_{\widehat\cG(x)}(v)$.
	Let  $(s_W^*,s_B^*)$ be an optimal strategy in $\widehat\cG(x)$, and consider a strategy $s:=(s_W^*,s_B)$, for some $s_B\in S_B$. By optimality of $s^*_W$ in $\widehat\cG(x)$, $v$ either reaches in $\widehat\cG_s(x)$ the self-loop $(w,w)$, or reaches another cycle $C$ of mean value (that is, $\sum_{e\in C}r(e)/|C|$) at least $\mu_{\widehat\cG(x)}(v)$. In both cases, since $y\ge\mu_{\widehat\cG(x)}$, $v$ reaches in $\widehat\cG_s(y)$ a cycle of mean value at least $\mu_{\widehat\cG(x)}(v)$, i.e., $\mu_{\widehat\cG(y)}(v)\ge\mu_{\widehat\cG(x)}(v)$. Now consider a strategy $s:=(s_W,s_B^*)$, for some $s_W\in S_W$. The optimality of $s^*_B$ in $\widehat\cG(x)$ implies that $v$ does not reach the self-loop $(w,w)$ in $\widehat\cG_s(x)$, since $x>\mu_{\widehat\cG(x)}(v)$. This implies that $\mu_{\widehat\cG(y)}(v)\le \mu_{\widehat\cG(x)}(v)$.
	
	(iii) Suppose that $\mu_{\widehat\cG(x)}(v)=x<y$. Let  $(s_W^*,s_B^*)$ be an optimal strategy in $\widehat\cG(x)$. In $\widehat\cG(y)$, let us consider a strategy $s:=(s_W^*,s_B)$, for some $s_B\in S_B$. By optimality of $s^*$ in $\widehat\cG(x)$, $v$ either reaches the self-loop $(w,w)$, or reaches in $\widehat\cG_s(x)$ another cycle of mean value at least $\mu_{\widehat\cG(x)}(v)$. This implies that, in $\widehat\cG(y)$, we either have $\mu_{\widehat\cG_s(y)}(v)=y>x$ or $\mu_{\widehat\cG_s(y)}(v) \ge \mu_{\widehat\cG(x)}(v)=x$. In both cases, we get $\mu_{\widehat\cG(y)}(v) \geq x$. Similarly, if $s:=(s_W,s_B^*)$, for some $s_W\in S_W$, then $v$ either reaches in $\widehat\cG_s(x)$ the self-loop $(w,w)$, or reaches another cycle of mean value at most $\mu_{\widehat\cG(x)}(v)=x<y$. This implies that, in $\widehat\cG(y)$, we either have $\mu_{\widehat\cG_s(y)}(v)=y$ or $\mu_{\widehat\cG_s(y)}(v) \le \mu_{\widehat\cG(x)}(v)<y$. In both cases, we get $\mu_{\widehat\cG(y)}(v) \leq y$.
	\qed

	\begin{corollary}\label{c1}
		For any position $v\in V$ of a parametrized BW-game $\widehat\cG(x)$ there is an interval $I(v):=[\lambda_1(v),\lambda_2(v)]$, such that 
		\begin{itemize}
			\item[(i)] $\mu_{\widehat\cG(x)}(v)=\lambda_2(v)$ if $x\ge \lambda_2(v)$;
			\item[(ii)] $\mu_{\widehat\cG(x)}(v)=\lambda_1(v)$ if $x\le \lambda_1(v)$;
			\item[(iii)] $\mu_{\widehat\cG(x)}(v)=x$ if $x\in I(v)$.
		\end{itemize}
	\end{corollary}
	\proof
	Consider a position $v\in V$. Then $\lambda_1(v):=\mu_{\widehat\cG(-R)}(v)$ and $\lambda_2(v):=\mu_{\widehat\cG(R)}(v)$ satisfy the conditions of the claim. Indeed, Lemma \ref{AB}-(i) and (ii) imply, respectively, claims (i) and (ii) of the corollary. Moreover, for any $x\in I(v)$, Claims (iii) and (iv) of Lemma \ref{AB} imply respectively that $\lambda_1(v)\le \mu_{\widehat\cG(x)}\le y$ and $y\le \mu_{\widehat\cG(x)}\le\lambda_2(v)$, and hence Claim (iii) of the corollary.
	\qed
	
	\begin{figure}
		\centering
		\begin{tikzpicture}[->,shorten >=1pt,auto,node distance=2cm,semithick,scale=0.8]
		\tikzstyle{every node}=[draw,shape=circle]
		
		\draw[step=1cm,gray,very thin] (0,0) grid (7.9,7.9);
		
		\draw[->,line width=1mm] (-0.5,0) -- (8,0);
		\draw[->,line width=1mm] (0,-0.5) -- (0,8);
		\draw[line width=1mm,blue] (0,2) -- (2.05,2) -- (4.05,4) -- (8,4);
		\draw[line width=1mm,red] (0,3) -- (2.95,3) -- (5.95,6) -- (8,6);
		\draw[line width=0.7mm,blue,dashed] (2,-.2) -- (2,2)  (4,-.2) -- (4,4);
		\draw[line width=0.7mm,red,dashed] (3,-.2) -- (3,3)  (6,-.2) -- (6,6);
		\node[draw=none,left] () at (0,7) {$\mu\left(\widehat{\mathcal{G}}_{(x)}\right)$};
		\node[draw=none,below] () at (7,0) {$x$};
		\node[draw=none,right] () at (8,6) {$\mu\left(\widehat{\mathcal{G}}_{(x)}(u)\right)$};
		\node[draw=none,right] () at (8,4) {$\mu\left(\widehat{\mathcal{G}}_{(x)}(v)\right)$};
		{\small
			\node[draw=none,below] () at (2,0) {$\lambda_1(v)$};
			\node[draw=none,below] () at (3,0) {$\lambda_1(u)$};
			\node[draw=none,below] () at (4,0) {$\lambda_2(v)$};
			\node[draw=none,below] () at (6,0) {$\lambda_2(u)$};
		}
		\end{tikzpicture}
		\caption{In this figure we illustrate Corollary~\ref{c1}. The two piecewise linear functions represent the value function of $\widehat{\mathcal{G}}(x)$ for initial positions $u$ and $v$, respectively, as functions of $x$. For the interval $I=[\lambda_1(u),\lambda_2(v)]$ we have $\{u,v\}\subseteq S^0(I)$, while for $I'=[\lambda_1(v),\lambda_1(u)]$ we have $u\in S^-(I')$ and $v\in S^0(I')$.  \label{fig:MFO-3}}
	\end{figure}
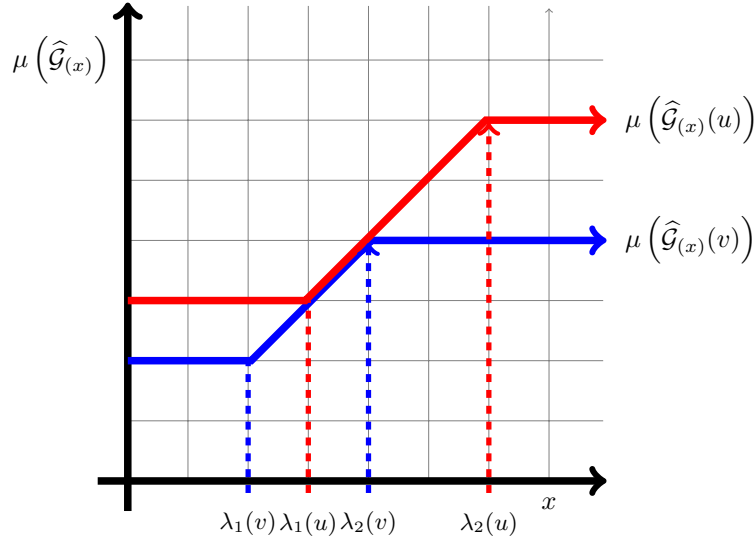

	Let $\widehat\cG(x)$ be a parametrized BW-game with a self-loop of reward $x$. By Corollary \ref{c1}, the end-points of the intervals $I(v)$, for $v\in V,$ partition the range $[-R,R]$ into a set of at most $2\nu(\widehat\cG)+1\le 2n+1$ intervals $\cI:=\cI(\widehat\cG(x))$, such that the "structure" of the game with respect to the dependence of the positions' values on $x$ is fixed over each interval. That is, for each $I=[\lambda_1(I),\lambda_2(I)]\in \cI$, there is a uniquely defined partition $S^-(I)\cup S^0(I)\cup S^+(I)=V$, such that
	\begin{itemize}
		\item $\mu_{\widehat\cG(x)}(v)=\lambda_2(v)$ for all $v\in S^-(I)$, where $\lambda_2(v)<\lambda_1(I)$; 
		\item $\mu_{\widehat\cG(x)}(v)=x$ for all $v\in S^0(I)$;
		\item $\mu_{\widehat\cG(x)}(v)=\lambda_1(v)$ for all $v\in S^+(I)$, where $\lambda_1(v)>\lambda_2(I)$.
	\end{itemize}
	Indeed, $S^0(I)$ is defined as the set $S$ such that $I=\bigcap_{v\in S}I(v)$; see Figure~\ref{fig:MFO-3} for an illustration. 
	
	It is not difficult to see that for each position $v$, both $\lambda_1(v)$ and $\lambda_2(v)$ are rational numbers with integer denominator not exceeding $n$ and integer numerator not exceeding $n R$.
	\begin{lemma}\label{lem:interval}
		Given a parametrized BW-game $\widehat\cG(x)$ and a position $v$, we can find $\lambda_1(v)$ and $\lambda_2(v)$ in time $O(n^5R\log(n R)).$
	\end{lemma}
	\proof
	The smallest possible value of $\mu_{\widehat\cG(x)}(v)$ is obtained for $x=-R$ and the largest for $x=R.$ After this we can do two binary searches to locate $\lambda_1$ and $\lambda_2$. Since in each step of the binary search we solve a BW-game with integer coefficients except for $x$ which is rational with denominator at most $n$, the required precision is only $\frac{1}{n^2}$. Consequently, we have to solve $O(\log(nR))$ BW-games and hence the claimed complexity bound follows by \cite{CR16}. 
	\qed
	
	We can now apply the above lemma for each position $v\in V$ and obtain a set of intervals $\{I(v)\mid v\in V\}$. From these we can obtain the set of intervals $\cI(\widehat\cG(x))$. 
	We shall call this procedure BW-FindIntervals$(\widehat\cG(x))$ which finds this set of intervals $\cI(\widehat\cG(x))$ together with the partitions $S^-(I)\cup S^0(I)\cup S^+(I)=V$, for $I\in\cI(\widehat\cG(x))$.
	
	
	
	In the algorithm, we will use the above results assuming that all random positions have the same value $x$, which we do not know exactly. By the above structure we will guess the interval $I$ containing $x$. Suppose that our guess is correct. It follows that the sets $S^+(I)$ and $S^{-}(I)$ contain no random positions and hence they provide a set of deterministic ergodic classes for which we obtain the values and optimal strategies by solving BW-games. On the other hand, the set $S^0(I)$ contains random positions for which the BW-game $\widehat{\cG[S^0(I)]}(x)$ has not enough information to obtain the optimal strategies. For this, we consider the parametrized BWR-game $\cG[S^0(I)](x)$, and find the interval of $x$-values for which this game is ergodic. By Lemmas~\ref{l2} and \ref{l3}, the optimal strategies obtained for such ergodic BWR-game will yield the optimal strategies for the corresponding set in $\cG$. 
	
	\begin{lemma}\label{l4}
		For a parametrized BWR-game $\cG(x)$, the set $I(\cG(x)):=\{x\in[-R,R]\mid \text{$\cG(x)$ is ergodic}\}$ forms a closed (possibly empty) interval in $[-R,R]$.
	\end{lemma}
	\proof
	Suppose that $\cG(x)$ is non-ergodic. Then either
	\begin{itemize}
		\item[(i)] there is a strategy $s_B^*\in S_B$ such that for all $s_W\in S_W$, there is position $v$ belonging to an absorbing class $C_{s_W}$ in the Markov chain $\cG_{(s_W,s_B^*)}(x)$ with value $\mu_{\cG_{(s_W,s_B^*)}(x)}<x$, or
		\item[(ii)] there is a strategy $s_W^*\in S_W$ such that for all $s_B\in S_B$, there is position $v$ belonging to an absorbing class $C_{s_B}$ in the Markov chain $\cG_{(s_W^*,s_B)}(x)$ with value $\mu_{\cG_{(s_W^*,s_B)}(x)}>x$.
	\end{itemize}
	In case (i), since $C_{s_W}$ does not include $w$, the same strategy $s_B^*$ guarantees that
	$\mu_{\cG_{(s_W,s_B^*)}(y)}<y$, for any $y>x$. In case (ii), the strategy $s_W^*$ guarantees that
	$\mu_{\cG_{(s_W^*,s_B)}(y)}>y$, for any $y<x$. Thus we conclude that the game $\cG(y)$ remains non-ergodic for either all $y\le x$ or all $y\ge x$.
	
	Suppose that there exist $\tau_1<\tau_2$ such that $\cG(\tau_1)$ and $\cG(\tau_2)$ are ergodic. Then the claim above implies that $\cG(\tau)$ is ergodic for all $\tau\in[\tau_1,\tau_2]$.
	\qed
	
	\begin{lemma}\label{l5}
		In a parametrized BWR-game $\cG(x)$, let $\tau_1\le\tau_2$ be two real numbers in $I(\cG(x))$, and $s_W^*\in S_W$ and $s_B^*\in S_B$ be optimal \Max\ and \Min\ strategies in the games $\cG(\tau_2)$ and $\cG(\tau_1)$, respectively. Then $(s_W^*,s_B^*)$ is a pair of optimal strategies in $\cG(x)$ for all $x\in [\tau_1, \tau_2]$.
	\end{lemma}
	\proof
	By the definition of $\cG(x)$ the position $w$ with the self-loop forms a single-vertex absorbing class with value $x$. Thus, for all $x\in[\tau_1,\tau_2]$, all other positions have the same value $x$, since $\cG(x)$ is an ergodic game. We claim that for all $v\in V$ and $s_W\in S_W$ we have $\mu_{\cG_{(s_W,s_B^*)}(\tau_1)}(v)\le \tau_1\le x$.
	To see this, consider a strategy $s=(s_W,s_B^*)$ for some $s_W\in S_W$.  In the Markov chain $\cG_s(\tau_1)$, any position $v$ in an absorbing class, that is not formed by the singleton $\{w\}$, has value $\mu_{\cG_s(\tau_1)}(v)\le \tau_1\le x$, and hence it has value $\mu_{\cG_s(x)}(v)\le \tau_1\le x$ in $\cG_s(x)$.  It follows that the value of any transient position $v$ in $\cG_s(x)$ is also at most $x$ since it is a convex combination of the values in the absorbing classes.
	
	Analogously, we obtain that for all $v\in V$ and $s_B\in S_B$: $\mu_{\cG_{(s_W^*,s_B)}(x)}(v)\ge x$.
	\qed
	
	To be able to compute efficiently $\cI(\cG(x))$ and the corresponding optimal strategies mentioned in the previous lemma, we need a procedure to find the top class of a given BWR-game. This is provided in the next lemma.
	\begin{lemma}\label{ll1}
		Let $\bar{\cG}:=\cG(\frac{p}{q})$ be a BWR-game obtained from the parametrized BWR-game $\cG(x)$, where $p,q$ are integers such that $p\le \sqrt{k}2^{k/2}D^kR$ and $q\le\sqrt{k}2^{k/2}D^k$.
		Then, we can determine both the bottom and top ergodic classes, and find a pair of strategies proving this in time $\poly(n)(2D)^{O(k)}R\log R$, using algorithm PUMP$(\bar\cG,\eps)$ with $\eps$ as in \raf{accur} .
	\end{lemma}
	\proof
	We first show how to find the top ergodic class (that is, the set of positions which have the highest value in $\bar \cG$).
	We apply Algorithm~\ref{top} called BWR-FindTop$(\bar\cG)$. It works by calling PUMP$(\bar\cG,\varepsilon)$ with $\eps$ as in \raf{accur}. If no contra-ergodic partition is produced then the game is ergodic and its solution has been already found by the selection of $\eps$, in view of Theorem~\ref{cor1}. Otherwise, any position $v \in V^-$ and hence in $\Cl'_B(V^-)$ has value less than the top value, which can be seen from the value equations (C1), and using induction as positions are added to $\Cl'_B(V^-)$. Thus if we remove all these positions we get a well-defined game $\cG':=\bar\cG[V\setminus\Cl_B'(V^-)]$ which includes the top class of $\bar\cG$. By induction, BWR-FindTop$(\cG')$ returns the top class $T\subseteq Y:=V\setminus\Cl_B'(V^-)$ in $\cG',$ and we claim that $T$ is the top class also in $\bar\cG$. To see this, note that by Proposition \ref{p1}, there are no black or random arcs from $T$ to $Y\setminus T$, and
	by the definition of $Y$, there are no black or random arcs from $T$ to $V\setminus Y$ either. Thus, if $s^*=s^*[T]$ is the situation returned by BWR-FindTop$(\cG')$, then $s_W^*$ guarantees for \Max\ in $\bar\cG$ the same value $y$ guaranteed by $s_W^*$ in $\cG'$. Furthermore, since $T$ is the top class in $\cG'$, there is a \Min\ strategy $\bar s_B=\bar s_B[Y]$ such that $\mu_{\bar\cG_{(s_W,\bar s_B)}} \le y$ for all \Max\ strategies $s_W=s_W[Y]$. By conditions (ii) and (iii) of the contra-ergodic partition, there is strategy $\hat s_B=\hat s_B[V^-]$ that forces \Max\ to stay in $V^-$ and ensures that $\mu_{\bar\cG_{(s_W[V^-],\hat s_B)}}(v)<y$ for all $v\in V^-$ and $s_W\in S_W$. It follows that the strategy $s_B\in S_B$ obtained by concatenating $\bar s_B[Y]$, $s_B(\Cl_B'(V^-))$ and $\hat s_B[V^-]$ satisfies $\mu_{\bar\cG_s}(v)\le y$ for all $v \in V$ and $s_W\in S_W$. This proves our claim. 
	
	Finding the bottom ergodic class is analogous and can be done by a similar procedure BWR-FindBottom$(\bar\cG)$.
	\qed
	\begin{algorithm}[htb]
		\caption{BWR-FindTop$(\cG)$}
		\label{top}
		\begin{algorithmic}[1]
			\REQUIRE A BWR-game $\cG=(G=(V,E),p,r)$
			\ENSURE A set $T\subseteq V$ such that for all $v\in V$, $\mu_{\cG}(v)=\max_{u\in V}{\mu_{\cG}(u)}$ and a situation $s^*$
			solving the game $\cG[T]$
			\STATE Set $\eps$ as in \raf{accur}
			\IF{PUMP$(\cG,\varepsilon)$ returns a contra-ergodic partiton $(V^+,V^0,V^-)$}
			\STATE FindTop$(\cG[V\setminus \Cl_B'(V^-)])$
			\ELSE
			\STATE Let $s^*$ be the situation returned by PUMP$(\cG,\varepsilon)$
			\RETURN $(V,s^*)$
			\ENDIF
		\end{algorithmic}
	\end{algorithm}
	
	\begin{lemma}\label{lem:ergodicity}
		Let us consider a parametrized BWR-game $\cG(x)$ and let $\cI(\cG(x))=[\tau_1,\tau_2]$. We can compute $\tau_1$, $\tau_2$, and the optimal strategies described in Lemma~\ref{l5} in time $\poly(n)(2D)^{O(k)}R\log^2 R$.
	\end{lemma}
	\proof
	We employ binary search, calling in each step the procedure BWR-FindTop$(\cG(x))$, defined in the previous lemma, with different guessed values of the parameter $x$. Suppose that we start the search on the interval $[\lambda_1,\lambda_2]$. If the top class does not include all the positions then the game is non-ergodic, and the procedure will return a position $v\in V$, and either a strategy $s_B\in S_B$ certifying that $\mu_{\cG(x)}(v)<x$ or a strategy $s_W\in S_W$ certifying that $\mu_{\cG(x)}(v)>x$. In the former case, we reduce the search into the interval $[ \lambda_1,x]$, and in the latter case, we reduce the search into the interval $[x, \lambda_2]$. Since by Lemma~\ref{l1-prob}, the precision needed is $(k^{1/2}2^{k/2}D^{k})^{-1}$, we need $O(\log R+k\log D)$ many search steps.
	\qed
	
	We call the procedure described in the above proof BWR-FindErgodicityInterval$(\cG(x))$.

	\subsection{Description of main algorithm}
	
	For a BWR-game $\cG$ and a parameter $x\in\RR$, define $\widehat\cG(x)$ to be the BW-game obtained from $\cG$ by
	replacing each random position $v\in V_R$ in $\cG$ with a terminal deterministic position (black or white, arbitrarily) with a local reward of value $x$ on the self-loop $(v,v)$.
	
	Our algorithm uses four auxiliary routines: BWR-FindErgodicityInterval$(\cG(x))$ and find BW-FindIntervals$(\cG(x))$ which we described above; BWR-SolveErgodic$(\cG)$ which solves a given ergodic BWR-game $\cG$ using the pumping algorithm as in Theorem~\ref{cor1}; and BW-Solve$(\cG)$ which solves a BW-game using e.g. \cite{GKK88,Pis99,ZP96}.

	For a  position $v \in V_R$, we define $\rank(v)=|\{\mu_{\cG}(u) \mid u \in V_R,~\mu_{\cG}(u)>\mu_{\cG}(v)\}|+1$.
	For each $v \in V_R$, our algorithm guesses  its rank as $g(v)$.
	We remark that  there are at most $k^k$ possible guesses.
	
	For each such guess $g:V_R \to [k]$, we call procedure BWR-Solve$(\cG,U,g,\ell,s^*[V\setminus U])$ with $U=V$ and $\ell=1$.  At any point in time, $U$ represents the set of positions for which strategies have not been fixed yet.
	
	This procedure keeps constructing complete situations for $\cG$ until it finds an optimal one or discovers that our current guess is not correct. Each time we check optimality by solving two MDPs using linear programming (see, e.g., \cite{MO70}).
	We will prove that for each guess $g$ the procedure will only construct $O(\nu(\cG)^k)$ complete situations. We will also prove that it always finds an optimal one if our guess is correct. 
	
	We now describe this procedure BWR-Solve$(\cdot)$.
	For an integer $\ell\in[k]$, define $U^\ell$ to be the set of positions obtained from $\cG$ by removing all positions in the black closure $\Cl_B(\{v\in V_R\mid g(v)>\ell\})$; these are positions for which the values are smaller than the value of the random positions at rank $\ell$, assuming that our guess is correct. We first form the game $\widehat {\cG[U^\ell]}(x)$. Then we find the set of intervals $\cI(\widehat{\cG[U^\ell]}(x))$ using the routine BW-FindIntervals$(\widehat{\cG[U^\ell]}(x))$ described above.
	Then for each such interval $I=[\lambda_1(I),\lambda_2(I)]$, we consider three subgames, defined by the sets $S^+(I)$, $S^0(I)$, and $S^-(I)$. By the definition of $S^+(I)$, the first subgame $\cG[S^+(I)]$ is a BW-game. Hence, the optimal strategy $s^*[S^+(I)]$ in $\cG[S^+(I)]$ can be obtained by calling BW-Solve$(\cG[S^+(I)])$. The positions in the second subgame $\cG[S^0(I)]$ have the same value $x$. Although we do not know what the exact value of $x$ is, we can find the interval of ergodicity of the BWR-game $\cG[S^0(I)](x)$ by calling procedure BWR-FindErgodicityInterval$(\cG[S^0(I)](x))$ (step \ref{s4-}). Once we determine this interval $[\tau_1,\tau_2]$, we solve the two ergodic games $\cG[S^0(I)](\tau_1)$ and $\cG[S^0(I)](\tau_2)$ using procedure BWR-IsErgodic$(\cdot)$, and then combine the strategies according to Lemma \ref{l5} to obtain an optimal situation for $\cG[S^0(I)]$.
	Finally, the rest of the game is solved by calling the procedure recursively with $\cG:=\cG[U\setminus (S^+(I)\cup S^0(I))]$ and $\ell:=\ell+1$.
	
	\begin{algorithm}[htb]
		\caption{BWR-Solve$(\cG,U,g,\ell,s^*[V\setminus U])$}
		\label{BWR-solve}
		\begin{algorithmic}[1]
			\REQUIRE A BWR-game $\cG=(G=(V=(V_B\cup V_W\cup V_R),E),p,r)$,  a set of positions $U\subseteq V$, a vector of rank guesses $g:V_R\to[k]$, an integer $\ell$, and a situation $s^*[V\setminus U]$ on the set of positions $V\setminus U$
			\IF{$V_R=\emptyset$}
			\STATE $s^*[U]:=$BW-Solve$(\cG[U])$ \label{s1-}
			\IF{$s^*$ is optimal in $\cG$}
			\STATE {\bf output} $s^*$ and {\bf halt}
			\ENDIF
			\ELSE
			\STATE $U^\ell:=U\setminus\Cl_B(\{v\in V_R\mid g(v)>\ell\})$
			\STATE $(\cI,S^-,S^0,S^+):=$BW-FindIntervals$(\widehat{\cG[U^\ell]}(x))$  \label{s2-}
			\FOR{each $I=[\lambda_1(I),\lambda_2(I)]\in \cI$}\label{s2.1}
			\STATE $s^*[S^+(I)]:=$BW-Solve$(\cG[S^+(I)])$  \label{s3-}
			\STATE $[\tau_1,\tau_2]:=$BWR-FindErgodicityInterval$(\cG[S^0(I)](x))$ \label{s4-}
			\STATE $s^1:=$BWR-SolveErgodic$(\cG[S^0(I)](\tau_1))$ \label{s5-}
			\STATE $s^2:=$BWR-SolveErgodic$(\cG[S^0(I)](\tau_2))$ \label{s6-}
			\STATE $s^*[S^0(I)]:=(s_W^2,s_B^1)$ \label{s7-}
			\IF {$U=(S^+(I)\cup S^0(I))$} 
			\IF{$s^*$ is optimal in $\cG$}
			\STATE {\bf output} $s^*$ and {\bf halt}
			\ENDIF
			\ELSE
			\STATE BWR-Solve$(\cG,U\setminus (S^+(I)\cup S^0(I)),g,\ell+1,s^*[(V\setminus U)\cup S^+(I)\cup S^0(I)])$ \label{s8-}
			\ENDIF
			\ENDFOR
			\ENDIF
		\end{algorithmic}
	\end{algorithm}
	
	The following lemma states that if the guess is correct,
	then the procedure actually solves the game.
	
	\begin{lemma}\label{l6}
		Let $\cG$ be a BWR-game. If procedure BWR-Solve$(\cG,U,g,\ell,s^*[V\setminus U])$ is called with $g(v)=\rank(v)$ for all $v\in V_R$, $U=V$ and $\ell=1$, then it returns an optimal situation $s^*$.
	\end{lemma}
	\proof 
	Suppose that $g$ is correct, i.e., $g(v)=\rank(v)$ for all $v\in V_R$. Let $d$ be a nonnegative integer such that $d\le \max_{v\in V_R}\rank(v)$. We prove by induction on $d$ that there is a path in the recursion tree (from the root to a node at depth $d-1$) through which  the algorithm finds correctly all the classes $S(\theta_i):=\{u\in V\mid \mu_{\cG}(u)=\theta_i\}$, for $\theta_i \ge \mu_{\cG}(v)$ for all $v\in V_R$ with $\rank(v)=d$, and that the obtained situation $s^*$ induces optimal situations in $\cG[\theta_i]$ for each such $i$. This claim together with Lemma \ref{l3} would prove the theorem.
	
	For $d=0$ there is nothing to prove. Suppose that the claim is correct up to $d=h-1$ and consider the path form the root of the recursion tree to a node $\cN$ verifying this. Consider the call to procedure BWR-Solve$(\cdot)$ at this node. Assume that $\mu_{\cG}(v)=x$ for all $v\in V_R$ with $\rank(v)=h$.
	Let $\cI$ be the set of intervals computed in step \ref{s2-}, and $I\in\cI$ be the interval for which $x\in I$; this interval will be eventually chosen in step \ref{s2.1}. Since all the positions in $\Cl_B(\{v\in V_R\mid g(v)>h\})$ have value less than $x$, Lemma \ref{l1}-(ii) implies that the values computed for all positions in the set $S^+(I)=\{v\in U^\ell\mid\mu_{\cG}(v)> x\}$ are correct.
	
	Furthermore, by definition of $S^0(I)$ the subgame $\cG[S^0(I)]$ is ergodic with value $x$, and hence $x$ belongs to the interval of ergodicity $[\tau_1,\tau_2]$ computed in step \ref{s4-}. By Lemma \ref{l5}, the situation $s^*[S^0(I)]$ computed in step \ref{s7-} is optimal in the subgame $\cG[S^0(I)]$.
	
	This completes the proof of the induction step.
	\qed

	
	\noindent{\bf Proof of Theorem~\ref{t-main}.} 
	The fact the the algorithm retunrs an optimal situation follows from the previous lemma.
	For the complexity analysis, note that the depth of the recursion tree is at most $k+1$. The number of intervals tried at each recursion level is at most $2\nu(\cG)+1$. The running time of the local work at each edge of the recursion tree is limited by $\poly(n)(2D)^{O(k)}R\log^2 R$ . We have at most $(2\nu(\cG)+1)^{k+1}$ such edges, and hence the claimed complexity follows since $\nu(\cG)\le n$.
	\qed
	
	Finally, we remark that our proof of Theorem \ref{t-main} actually gives the stronger bound of $(\nu Dk)^{O(k)}R\cdot\poly(n,\log R)$ for the running time of the algorithm. The definition of $\nu(G)$ implies the following observation.
	
	\begin{observation}\label{nu}
		For a stochastic terminal payoff game with $t$ terminals, 
		$\nu(\cG)\le t+1$.
	\end{observation}
	Thus, we obtain the following result. 
	\begin{corollary}\label{cor:c1}
		Algorithm~\ref{BWR-solve} solves any stochastic terminal payoff game with $t$ terminals in time $(t Dk)^{O(k)}R\cdot \poly(n,\log R)$, and any simple stochastic game in time $k^{O(k)}\poly(n)$. 
	\end{corollary}

\section*{Acknowledgements} The first author thanks the National Science Foundation for partial support (Grant IIS-1161476). The third author was partially supported by Grant-in-Aid for Scientific Research from the Ministry of Education, Culture, Sports, Science and Technology of Japan.
The authors are also grateful for the support received from the Research-in-Pairs Program of the Mathematisches Forschungsinstitut Oberwolfach. 
\newcommand{\etalchar}[1]{$^{#1}$}


\end{document}